\documentclass[english,a4paper,11pt,headings=standardclasses]{scrartcl}
\usepackage{iftex}
\ifPDFTeX
   \usepackage[utf8]{inputenc}
   \usepackage[T1]{fontenc}
   \usepackage{lmodern}
\else
   \ifXeTeX
     \usepackage{fontspec}
   \else 
     \usepackage{luatextra}
   \fi
   \defaultfontfeatures{Ligatures=TeX}
\fi

\usepackage[UKenglish]{babel}

\usepackage{amssymb}
\usepackage{mathtools}
\usepackage{bm} 
\usepackage{upgreek} 
\usepackage{siunitx} 
\DeclareSIUnit\atm{atm}
\DeclareSIUnit\bases{bp}
\usepackage{chemformula} 

\usepackage{color}

\usepackage{graphicx} 
\graphicspath{{./images/}}
\usepackage{float} 
\usepackage{subcaption} 

\usepackage{tikz}
\usetikzlibrary{decorations.pathmorphing}
\usetikzlibrary{babel}
\tikzstyle{spring}=[thick,decorate,decoration={zigzag, pre length=6, post length=6,segment length=6}]

\usepackage[colorlinks=true]{hyperref} 
\hypersetup{linktocpage=true,citecolor=blue,linkcolor=blue,urlcolor=blue} 
\usepackage{cleveref} 
\numberwithin{figure}{section} 
\numberwithin{equation}{section}
\numberwithin{table}{section} 
\usepackage{relsec} 

\usepackage{booktabs} 

\usepackage[autostyle,spanish=spanish]{csquotes} 
\usepackage[square,numbers]{natbib}
\usepackage{listings}
\usepackage{abstract}
\usepackage{afterpage} 
\usepackage{fancyhdr} 
\usepackage{subfiles}
\usepackage{tensor}
\usepackage{hyperref}
\usepackage{cancel}
\usepackage{commath}
\usepackage{multirow}

\usepackage[top=2.75cm, bottom=2.75cm, left=3.00cm, right=2.50cm]{geometry}
\usepackage{afterpage}
\usepackage[normalem]{ulem}
\useunder{\uline}{\ul}{}

\renewcommand{\d}[1]{\ensuremath{\operatorname{d}\!{#1}}}
\DeclareSIUnit\parsec{pc}
\DeclareSIUnit\year{yr}
\cleardoublepage

\def\IN{\mathbb{N}}

\def\bi{\begin{itemize}}
\def\ei{\end{itemize}}

\def\a{{\alpha}}
\def\b{{\beta}}

\def\g{{\gamma}}

\def\be{\begin{equation}}
\def\ee{\end{equation}}
\def\bea{\begin{eqnarray}}
\def\eea{\end{eqnarray}}

\newfloat{lstfloat}{htbp}{lop}
\floatname{lstfloat}{Listing}

\begin{document}
\begin{titlepage}

    \newcommand{\HRule}{\rule{\linewidth}{1.25mm}} 
    
    \vspace{2cm}
    \center 
    
    \HRule \\[0.4cm]
    { \huge \bfseries Quantum corrections to Einstein's equations }\\[0.03cm] 
    \HRule \\[2cm]

    \textsc{\Large Master Thesis Memory}\\[1.75cm] 
    
    \begin{minipage}{0.9\textwidth}
        \begin{center} \Large
        \emph{Authors:}\\
        Eduardo \textsc{Velasco-Aja} and Jesús \textsc{Anero}\\\href{mailto:eduardovelascoaja@gmail.es}{\small eduardovelascoaja@gmail.es} \hspace{2cm}\href{mailto:jesusanero@uam.es}{\small jesusanero@uam.es}
        \end{center}
    \end{minipage}\\[1.75cm]
        
        
        \textsc{\Large Dpto.~de Física Teórica }\\[0.3cm]
        \textsc{\Large Universidad Autónoma de Madrid }\\[0.6cm] 
        \textsc{\Large Instituto de Física Teórica }\\[0.3cm]
        \textsc{\Large UAM/CSIC }\\[1.75cm] 

    
    
    
    
    {\large September 2022 } 
    

    \vfill 
    
    \end{titlepage}


\begin{abstract}
    In this master thesis, the Frobenius power series expansion method is used to find spherically
    symmetric and static vacuum solutions to quadratic and cubic gravitational actions, representing quantum corrections to the Einstein-Hilbert action. After
    a motivation to the topic and an introduction to the formal aspects, the power series
    solutions are presented. After recovering the results for the quadratic action of \footnote{H. Lü, A. Perkins and K.S. Stelle, 'Spherically symmetric solutions in higher-derivative gravity', \href{https://doi.org/10.1103/PhysRevD.92.124019}{10.1103/PhysRevD.92.124019} (2015). }\footnote{E. Alvarez, J. Anero and R. Santos-Garcia, 'Structural stability of spherical horizons', \href{https://doi.org/10.1088/1361-6382/ac25e5}{10.1088/1361-6382/ac25e5} (2010).}, we found that when the Weyl cubic operator is present, the \((2,2)\) family of solutions is still present while the Schwarzschid-de Sitter-like \((1,-1)\) is not.
    Lastly, we briefly discuss some of the solutions’ properties.
\end{abstract}
\newpage
\dynsection*{Foreword}\label{sec:foreword}
This e-print corresponds to the master thesis of one of the authors (E.V.), performed under the supervision of professor Enrique Álvarez Vázquez of UAM-IFT for the 2021/2022 academic year.

The role of J.A. was pivotal to understand the work previously done in \cite{Alvarez2020}. This included extended help with the programming and infinite patience for code-debugging, which ended up extending to manuscript redaction and revision.

The work here presented constituted the basic study for some of the series solutions to cubic actions that were later included in \cite{Alvarez2022}.
Lastly, this e-print serves as a broader study on power series static and spherically symmetric solutions to gravitational actions with quadratic and cubic curvature operators.
\newpage
\tableofcontents
\newpage
\clearpage
\dynsection{Introduction and summary of results.}\label{sec:1}
Effective field theories date back to Wilson \cite{Wilson} and constitute the groundings for many state-of-the-art approaches to particle physics.
\\
In the context of gravitation \cite{Buchbinder1992,Vilkovisky1992}, one considers diffeomorphism invariance as the \textbf{fundamental symmety} of the action. 

Then, it seems unavoidable to adopt the view that the Einstein-Hilbert Lagrangian (linear in curvature) is but the lowest energy approximation (deep infrared) of a more general lagrangian involving higher order Diff-invariant curvature operators (perhaps all of them).

We can write this sum of operators as,  
\begin{equation}
    S=\int d^4x\sqrt{\abs{g}}\left\{-\Lambda+C_1\,R+C_2\,R^2+C_3\, \tensor{R}{_\mu_\nu}\tensor{R}{^\mu^\nu}+C_4\,\tensor{R}{_\mu_\nu_\zeta_\xi}\tensor{R}{^\mu^\nu^\zeta^\xi}+\cdots\right\}.\label{1}
\end{equation}
In \cref{1}, the first term is the Cosmological Constant, whose smallness remains to be properly understood. 
This is the \underline{Cosmological Constant problem} see \cite{Weinber1989,Nobbenhuis2006} for an overview.

The next term constitutes the Einstein-Hilbert action. Variation of these first two terms in the action gives the equations of motion (EoMs) of General Relativity.

Next, we have the quadratic terms in the Riemann tensor or any of its contractions. These are preceded by the coefficients \(C_2,\,C_3,\,C_4\). 
In four dimensions we have the Gauss-Bonnet topological invariant (\cref{sec:A}) which means that, in four dimensions, there are only two independent quadratic-curvature integrated invariants.

Thus we can rewrite the action \eqref{1} as, 
\begin{equation}
    S=\int d^4x\sqrt{\abs{g}}\left\{-\Lambda-\gamma\,R-2\alpha\,R^2+\left(\beta+\frac{2\alpha}{3}\right) \tensor{R}{_\mu_\nu}\tensor{R}{^\mu^\nu}+\cdots\right\},\label{3}
\end{equation}
where, in order to recover Newtonian Gravity, \(\gamma=1/16\pi G:=1/2\kappa^2\), with \(G\) Newton's Gravitational Constant. The dots represent all the higher order operators in curvature.

It is important to note that, on dimensional grounds, \( \alpha\) and \(\beta\) are dimensionless, whereas any higher-order operators will be preceded by dimensionful coefficients, meaning from quadratic operators on, the operators are irrelevant in the Renormalization Group sense.

The objective of this work is to consider the solutions to the EoMs derived from an action directly inspired by the Effective Field Theory approach.
That way, we explore the fate of the spherically symmetric static solutions to the EoMs when higher-order curvature operators, corresponding to quantum corrections to GR are present.

The question is then, how to truncate the potentially infinite sum of operators in a way such, that the task of studying the EoMs is achievable and most importantly, physically driven.

We can answer this question bearing in mind the work done on the Renormalization of pure Einstein-Hilbert gravity to one loop by 't Hooft and Veltman \cite{Hooft} and to two loops by Goroff and Sagnotti \cite{Goroff1985}. 
The former demonstrated that the counterterm of pure gravity to one loop is a combination of \(R^2\) and \( \tensor{R}{^2_\mu_\nu}\) both vanishing on-shell. 
The latter showed that to two loops pure gravity diverges on-shell. 
What is more, the divergent term was shown to be proportional in vacuum to the Weyl cubed operator.
\\
Thus, the action we will explore is 
\begin{equation}
S=\int d^4 x \sqrt{|g|} \, \Big\{-\Lambda-\gamma \, R-2 \alpha \, \tensor{R}{_\mu_\nu}^2+\left(\beta+\frac{2}{3} \alpha \right)\,R^2+\omega\kappa^2 \tensor{C}{_\mu_\nu_\zeta_\xi}\tensor{C}{^\zeta^\xi^\rho^\sigma}\tensor{C}{_\rho_\sigma^\mu^\nu}\Big\},\label{Act}
\end{equation}
where, \(\tensor{C}{_\mu_\nu_\zeta_\xi}\) is the Weyl tensor defined as a Riemann curvature tensor to which all contractions have been substracted see \cref{sec:A}.

The Weyl cubed term is preceded by a \(\omega\kappa^2\) coefficient such that \(\omega\) is dimensionless.
\\
The plan is as follows; in \cref{sec:2}, we will discuss and derive the EoMs for the action \eqref{Act}. 
For this, in \cref{sec:2.1} we first use traditional variational methods, obtaining the Diff-invariant equations of motion.

Still and all, we will only consider solutions corresponding to \textbf{spherically symmetric static metrics} written on the \textit{Schwarzschild-Gauge}\footnote{As discussed in \cref{sec:D} the Nariai Metric, has spherical symmetry but cannot be cast into this gauge. Hence we are not even considering the entirety of spherically symmetric solutions but a subset of them.}.

Such metrics can be written, see \cref{sec:D}, as,
\begin{equation}
    \d s^2=B(r)\,\d t^2-A(r)\d r^2-r^2\left(\d \theta^2+\sin^2 (\theta)\d \phi^2\right)=\tensor{\mathbf{g}}{_\mu_\nu}\d x^\mu\d x^\nu.\label{metans}
\end{equation}
The high degree of symmetry in \eqref{metans} allows to alternatively, derive the EoMs for \(A(r), \;B(r)\) by first imposing \eqref{metans} in the action to later derive the EoMs. 
We will name this approach \textit{Symmetric Criticality}. 
The mathematical basis for this method is due to Palais \cite{Palais}, and was recently revisited in \cite{Fels2002}.
This approach was first introduced in the context of GR by Weyl \cite{Weyl1952} and later applied to a wider variety of actions by Deser and Tekin \cite{Deser2003}.

Once the EoMs for \(A(r), \; B(r)\) are obtained, given the lack of closed-form solutions we will try to find power series solutions for them.
This approach is in no way new \cite{Stelle1978,Stelle2015, Alvarez2020}, however, it remains a powerful tool to obtain solutions up to any desired order even when a closed form for the series is not known. 
For example, the first coefficients on the power series may give information about the topological properties of space-time, such as the presence of Horizons \cite{Stelle2015, Stelle2015-2, Kehagias2015, Alvarez2020, Holdom2002, Holdom2016, Holdom2019, Pravda2021}.

We will start our discussion on power series solutions in \cref{sec:3}. Here we discuss the approach taken when looking for power series solutions. In section \cref{sec:3.3} we go over some examples.

After this, we apply the power series Ansatz to the functions \(A(r)\) and \(B(r)\) in \cref{metans}. 
We will assume they admit a series expansion of the form
\begin{align}
    A(r)&:=r^s\left(a_s+a_{s+1}\, r+a_{s+2}\, r^2 +\cdots\right)\label{Aans},\\
    B(r)&:=b_t r^t\left(1+b_{t+1}\, r+b_{t+2}\, r^2 +\cdots\right)\label{Bans},
\end{align}
where \((s,t)\in \mathbb{Z}\), \(a_s,b_t\neq0\). 
We write \eqref{Bans} in such form because the coefficient \(b_t\) can be re-absorbed by a time coordinate re-scaling.
In \cref{sec:3.2} we study the families \((s,t)\) for which the EoMs can be solved depending on the values of the parameters \(\alpha, \, \beta, \, \mbox{and } \omega\).
\\
Then, in \cref{sec:4} we discuss the power series solutions found. This work was started for the \(\omega=0\) case in \cite{Stelle1978, Stelle2015, Alvarez2020}, nonetheless, only  the \(\alpha,\beta\neq 0\) case was fully covered. 
To our knowledge, the \(\omega\neq0\) scenario has never been explored before.
We will find that even for \(\omega=0\), allowing particular values of \(\alpha, \beta\) yields what, to our knowledge, are new series solutions to the EoMs, previously only partially discussed in \cite{Stelle2015}. These new solutions contain, in certain adequate cases, some of the closed solutions for quadratic gravity models \cite{Pravda2021, Kehagias2015, Nelson2010}.

The interest of these power series solutions goes presumably beyond the Quantum Gravity realm. 
It could happen that the \((2,2)\) solution, or for that matter, any other Non-Schwarzschild solution might present characteristic features that distinguish them from the Schwarzschild-de Sitter solution.
Some authors defend that such differences could potentially be measurable by gravitational radiation experiments observing black-hole mergers \cite{Holdom2016, Holdom2020, Cardoso2016}. 

Although interesting, we hold the more conservative opinion that more work needs to be done before a conclusion can be drawn. Hence, we will motivate our study by its relation to EFTs and its intrinsic interest.
\dynsection{Equations of Motion.}\label{sec:2}
\leveldown{}
\dynsection{Variational Method.}\label{sec:2.1}
In order to obtain the EoMs for the full action in \eqref{Act}, we can split the variation into three parts,
\begin{equation}
    S=S_{\tiny\mbox{E-H}}\left(\gamma,\Lambda\right)+S_{\tiny\mbox{Quad}}\left(\alpha,\beta\right)+S_{\tiny\mbox{Weyl}^3}\left(\omega\kappa^2\right),
\end{equation}
where, 
\begin{align}
    S_{\tiny\mbox{E-H}}\left(\gamma,\Lambda\right)&:=S=\int d^4 x \sqrt{|g|} \, \left(-\Lambda-\gamma \, R\right),\label{act1}\\
    S_{\tiny\mbox{Quad}}\left(\alpha,\beta\right)&:=S=\int d^4 x \sqrt{|g|} \, \left(-2 \alpha \, \tensor{R}{_\mu_\nu}^2+\left(\beta+\frac{2}{3} \alpha \right)\,R^2\right),\label{act2}\\
    S_{\tiny\mbox{Weyl}^3}\left(\omega\kappa^2\right)&:=S=\int d^4 x \sqrt{|g|} \,  \left(\omega\kappa^2\, \tensor{C}{_\mu_\nu_\zeta_\xi}\tensor{C}{^\zeta^\xi^\rho^\sigma}\tensor{C}{_\rho_\sigma^\mu^\nu}\right).\label{act3}
\end{align}
Then, the variation of the total action will, by linearity, be the sum of the variations, 
\begin{equation}
    \tensor{H}{_\mu_\nu}:=\frac{1}{\sqrt{\abs{g}}}\frac{\delta\quad}{\delta \tensor{g}{^\mu^\nu}}\,S=\frac{1}{\sqrt{\abs{g}}}\frac{\delta\quad}{\delta \tensor{g}{^\mu^\nu}}\,\left(S_{\tiny\mbox{E-H}}\left(\gamma,\Lambda\right)+S_{\tiny\mbox{Quad}}\left(\alpha,\beta\right)\right)+\frac{1}{\sqrt{\abs{g}}}\frac{\delta\quad}{\delta \tensor{g}{^\mu^\nu}}\,S_{\tiny\mbox{Weyl}^3}\left(\omega\kappa^2\right),
\end{equation}
And we can exploit the fact that the EoMs for the E-H with the quadratic terms was already given in \cite{Alvarez2020,Stelle1978,Stelle2015,Nelson2010,Pravda2021} among many other references.

To find the EoMs for \eqref{Act}, it only remains to vary the Weyl cubic part.

The EoMs for \(S_{\tiny\mbox{Weyl}^3}\) correspond to,
\begin{align}
&\frac{12}{(n-1)(n-2)}R C_{\mu\a\b\lambda}C_{\nu}^{~\a\b\lambda}-\frac{12}{n-2}R^{\a\b}F_{\mu\a\nu\b}-\nonumber\\
&-\frac{6}{(n-1)(n-2)}R_{\mu\nu}C^2+3R_{\mu}^{~\a\b\lambda}F_{\nu\a\b\lambda}-6C_{\mu}^{~\a\b\lambda}F_{\nu\a\b\lambda}+\frac{1}{2}g_{\mu\nu}C^3-\nonumber\\
&-\frac{12}{n-2}\nabla_\tau\nabla_\mu\left(C_{\nu\a\b\lambda}C^{\tau\a\b\lambda}\right)+\frac{6}{(n-1)(n-2)}\left(\nabla_\mu\nabla_\nu-g_{\mu\nu}\Box\right)C^2+\nonumber\\
&+\frac{6}{n-2}\Box C_{\mu\a\b\lambda}C_{\nu}^{~\a\b\lambda}+\frac{6}{n-2}g_{\mu\nu}\nabla_\rho\nabla_\sigma C^{\rho\a\b\lambda}C^{\sigma}_{~\a\b\lambda}-6\nabla^\a\nabla^\b F_{\mu\a\nu\b}=0,\label{2.1}  
\end{align}
where we have defined,
\begin{align}
F_{\mu\nu\rho\sigma}&:=  C_{\mu\nu\a\b}C^{\a\b}_{~~\rho\sigma},\nonumber\\
C^3&:=  C_{\mu\nu\a\b}C^{\a\b\rho\sigma}C_{\rho\sigma}^{~~\mu\nu},\nonumber\\
C^2&:=  C_{\mu\nu\a\b}C^{\mu\nu\a\b}=F_{\mu\nu}^{~~\mu\nu}.
\end{align}
In 4 dimensions, we can use the Lanczos identity \cite{Lanczos1938}
\begin{equation}
    C_{\mu\a\b\lambda}C_{\nu}^{~\a\b\lambda} =\frac{1}{4}g_{\mu\nu}C^2, 
\end{equation}
and the explicit form of the Weyl tensor \eqref{A.7}, to write the EoMs for \( S_{\tiny\mbox{Weyl}^3}\), as 
\begin{align}
H^{\tiny\mbox{Weyl}^3}_{\mu\nu}&=\omega\kappa^2\Big[\frac{1}{2}g_{\mu\nu}C^3+\frac{1}{4}\left(R\,g_{\mu\nu}-R_{\mu\nu}\right)C^2-\frac{1}{2}\left(\nabla_\mu\nabla_\nu-g_{\mu\nu}\Box\right)C^2-\nonumber\\
&-3R^{\a\b}F_{\mu\a\nu\b}-3C_{\mu}^{~\a\b\lambda}F_{\nu\a\b\lambda}-6\nabla^\a\nabla^\b F_{\mu\a\nu\b}\Big]=0.
\end{align}
Thus, if we combine the EoMs for \(S_{\tiny\mbox{Weyl}^3}\) with the ones for \(S_{\tiny\mbox{E-H}}+S_{\tiny\mbox{Quad}}\), we arrive at the full EoMs for \eqref{Act}, 
\begin{align}
    H_{\mu\nu}&=\g\left(R_{\mu\nu}-\frac{1}{2}R\,g_{\mu\nu}\right)-2\left(\b+\frac{2\a}{3}\right)R\left(R_{\mu\nu}-\frac{1}{4}R\,g_{\mu\nu}\right)+4\a R^{\rho\sigma}\left(R_{\mu\rho\nu\sigma}-\frac{1}{4}R_{\rho\sigma}\,g_{\mu\nu}\right)+\nonumber\\
    &+\frac{2}{3}\left(\a-3\b\right)\left(g_{\mu\nu}{\Box}-\nabla_\mu\nabla_\nu\right)R+2\a \Box\left(R_{\mu\nu}-\frac{1}{2}R\,g_{\mu\nu}\right)-\frac{1}{2}g_{\mu\nu}\Lambda+\nonumber\\
    &+\omega\kappa^2\Big[\frac{1}{2}g_{\mu\nu}C^3+\frac{1}{4}\left(R\,g_{\mu\nu}-R_{\mu\nu}\right)C^2-\frac{1}{2}\left(\nabla_\mu\nabla_\nu-g_{\mu\nu}\Box\right)C^2-\nonumber\\
    &-3R^{\a\b}F_{\mu\a\nu\b}-3C_{\mu}^{~\a\b\lambda}F_{\nu\a\b\lambda}-6\nabla^\a\nabla^\b F_{\mu\a\nu\b}\Big]=0.\label{Eoms}
    \end{align}
If we solve \cref{Eoms} for the diagonal \textit{Schwarzschild Gauged} metric of \cref{metans}, we find the EoMs to be diagonal. Additionally, from spherical symmetry we have,
\begin{equation}
    H_{\phi\phi}=\sin^2\left(\theta\right)H_{\theta\theta}.\label{cond.1}
\end{equation}
From Diff-invariance we have the Bianchi identity, which gives the additional constraint, 
\begin{equation}
    \nabla_\mu H^{\mu\nu}=0.\label{cond.2}
\end{equation}
That is, using \eqref{cond.1} and \eqref{cond.2}, we have that only two out of the four non-vanishing components of $H_{\mu\nu}$  will be independent.
We will consider, as the two independent EoMs, the components, $H_{tt}$ and $H_{rr}$,
\begin{align}
    H_{tt}&=0,\label{eom.1.1}\\
    H_{rr}&=0.\label{eom.2.1}
\end{align}
The full expressions for \eqref{eom.1.1} and \eqref{eom.2.1}, when substituting the metric of \cref{metans} are given in \cref{sec:B}.

\dynsection{Symmetric Criticality.}\label{sec:2.2}
The EoMs in \eqref{Eoms} were derived in a diffeomorphism-invariant way\footnote{Since they are obtained as a variation with respect to the metric, they will inherit all symmetries the metric has and will preserve its invariance under changes of coordinates.}.
However, from the get-go, we knew, we are not concerned with the most generally written, Diff-invariant metric. 
In \cref{sec:1} we stated our interest in a very particular kind of metrics. 
Static, spherically symmetric, and with the \textit{Schwarzschild Gauge}.
\\
With this in mind, one might ask under which circumstances, if any, we could impose the desired symmetries of the metric beforehand to achieve a certain degree of simplification when deriving the EoMs.

This idea was first used by Weyl to derive Schwarzschild's solution \cite{Weyl1952}, and was later applied to more general actions by Deser and Tekin \cite{Deser2003}, who not only provided several examples of the method but also stressed its limitations, given by the Principle of Symmetric Criticality of Palais \cite{Palais, Fels2002}.

We will refer to this approach of first imposing the symmetries to the action to then derive the EoMs as \textit{Symmetric Criticality}.
Let us highlight the basis of this approach.
\\
Given a gravitational Diff-invariant action, 
\begin{equation}
    S=\int d^4 x\, \sqrt{\abs{g}}\; \mathcal{L}\left(x^\mu,\tensor{g}{_\mu_\nu},\partial\tensor{g}{_\mu_\nu},\partial^2\tensor{g}{_\mu_\nu}\right),\label{2.2}
\end{equation}
we can calculate using \cref{metans}
\begin{equation}
    \left.\mathcal{L}\left(x^\mu,\tensor{g}{_\mu_\nu},\partial\tensor{g}{_\mu_\nu},\partial^2\tensor{g}{_\mu_\nu}\right)\right\rvert_{g\to \mathbf{g}}=\mathcal{L}\left(x^\mu,A(r),B(r),\partial A(r),\partial B(r),\partial^2 A(r),\partial^2 B(r)\right),
\end{equation}
from staticity, spherical symmetry and the fact that the only dynamical variable of \(x^\mu\) is \(r\). We can integrate out  \(\{t,\theta,\phi\}\) in \cref{2.2}, which becomes,
\begin{equation}
    S=\int dr \sqrt{\abs{g}}\; \tilde{\mathcal{L}}\left(r,A(r),B(r), A'(r),\,B'(r),\,A''(r),\, B''(r)\right),\label{2.3}
\end{equation}
where the comma denotes derivation with respect to \(r\), i.e., \((\cdot)':=\partial_r\;(\cdot)\), and
\begin{equation}
    \sqrt{\abs{g}}=\sqrt{A(r)B(r)}\;r\,\sin (\theta).
\end{equation}
Then, the EoMs can be obtained as the two Euler Lagrange equations for \(A(r)\) and \(B(r)\)
\begin{align}
    \frac{\partial \sqrt{\abs{g}}\mathcal{L}}{\partial A(r)}-\frac{d}{dr}\frac{\partial \sqrt{\abs{g}}\mathcal{L}}{\partial A'(r)}+\frac{d^2}{d\,r^2}\frac{\partial \sqrt{\abs{g}}\mathcal{L}}{\partial A''(r)}&=0,\label{eom.1.2}\\
    \frac{\partial \sqrt{\abs{g}}\mathcal{L}}{\partial B(r)}-\frac{d}{dr}\frac{\partial \sqrt{\abs{g}}\mathcal{L}}{\partial B'(r)}+\frac{d^2}{d\,r^2}\frac{\partial \sqrt{\abs{g}}\mathcal{L}}{\partial B''(r)}&=0.\label{eom.2.2}
\end{align} 
\textit{Symmetric Criticality} does not only provide us with a consistency check for the EoMs derived in the full Diff-invariant formalism. 
Most importantly, it has the remarkable feature of making the structure of the action simpler than its Diff-invariant counterpart.
\levelup{}
\dynsection{Solutions to non-linear ODE systems.}\label{sec:3}
In this section we present the main ideas behind the power series solution to the non-linear Ordinary Differential Equation (ODE) system given by \eqref{B.1.a} and \eqref{B.2}.

In the context of second-order, linear ODEs the method of finding solutions with the form of a power series is often referred to as \textbf{Frobenius analysis}. 
Nevertheless, the idea of considering a power series solution is more general and can be applied to a system of non-linear ordinary differential equations.
Even if one should bear in mind the different contexts of the two, we will, for convenience,  refer to the power series solutions found for \eqref{B.1.a} and \eqref{B.2} as \textit{Frobenius solutions}.

In section \ref{sec:3.1} we go over the approach taken to solve equations \eqref{B.1.a} and \eqref{B.2} using power series solutions. 
After this, in \cref{sec:3.3} we consider the expansion of well known solutions to highlight some properties of power series solution.
Lastly,  in \cref{sec:3.2} we analize the equivalent of the indicial equation in a Frobenius analysis.

\leveldown{}
\dynsection{Power series analysis.}\label{sec:3.1}
To find solutions, we will truncate the power series ansatz in \cref{metans,Aans,Bans} at a given power and substitute it into the EoMs.
This results in the set of two EoMs being exact up to a particular order in \(r\). 

Then, with the exact coefficients in the EoMs, we obtain a system of Equations for the coefficients of the truncated functions \(A(r),\,B(r)\), i.e., \(\{a_s,\,a_{s+1},b_t,\,b_{t+1},\cdots\}\).
This system is solved to find a set of relations between the coefficients in each case. 
The number of parameters that are not fixed by the system constitutes the free parameters of the solution. The counting of the parameters is done without taking into account \(\Lambda\), nor \(b_t\) \footnote{This parameter can be set to 1 by a time re-scaling}. 
In certain cases the number of parameters does not remain fixed as with each new order for \(A(r) \) and \(B(r)\) two parameters are introduced but the two coefficients of the EoMs are not independent. 

Checking a particular ansatz works for arbitrarily large powers is complicated. 
Since solutions are rejected when the EoMs require any of the by assumption non-vanishing parameters \(\{\alpha, \,\beta,\,\gamma,\,\omega\}\) go to zero.
Once we find coefficients depending on all non-vanishing parameters of the action consistently, we will consider a particular \((s,t)\) family to be a \textit{consistent solution}.

\dynsection{Some already known power series.}\label{sec:3.3}
Before considering power series solutions whose sum we might not be able to obtain, let us first warm-up considering some general properties of the series expansion of already known solutions.

Let us start with the Schwarzschild-de Sitter solution;
\begin{equation}
    \d s^2=\left(1-\frac{R_s}{r}-\frac{\Lambda}{6\gamma}r^2 \right)\d t^2-\left(1-\frac{R_s}{r}-\frac{\Lambda}{6\gamma}r^2 \right)^{-1}\d r^2-r^2\d \Omega_2,
\end{equation}
which can easily be converted into pure Schwarzschild, or pure  (Anti) de Sitter letting \(\Lambda\to 0\) or \(R_s\to0\) respectively.
\\
The temporal part, \(B(r)\), has a trivial expansion around 0, while the radial part, \(A(r)\), can be expanded, defining, \(r_\Lambda^2:=6\gamma/\Lambda\), as 
\begin{equation}
    \frac{1}{1-\frac{R_s}{r}-\frac{r^2}{r_\Lambda^2}}=-\frac{r}{R_s}-\frac{r^2}{R_s^2}-\frac{r^3}{R_s^3}+\frac{r^4}{R_s^4}\left(\frac{R_s^2}{r_{\Lambda }^2}-1\right)+\frac{r^5}{R_s^5} \left(\frac{2 R_s^2}{r_{\Lambda }^2}-1\right)+\mathcal{O}\left(r^{6}\right), \label{s.1}
\end{equation}
in the cases where one of the two parameters vanishes, we formaly have 
\begin{align}
    &\bullet R_s=0\longrightarrow\;\frac{1}{1-\frac{r^2}{r_\Lambda^2}}=\sum_{n=0}^{\infty}\left(\frac{r}{r_\Lambda}\right)^{2n},\label{s.2}\\
    &\bullet r_\Lambda=0\longrightarrow\;\frac{1}{1-\frac{R_s}{r}}=-\frac{r}{R_s}\sum_{n=0}^{\infty}\left(\frac{r}{R_S}\right)^{n}.\label{s.3}
\end{align}
If we found a terminating series as a solution, we would have an exact solution to \eqref{eom.1.1} and \eqref{eom.2.1}, but the converse is not necessarily true. 
\cref{s.1,s.2,s.3} above, show certain exact solutions\footnote{The actions to which SdS is a solution are presented in \cref{sec:4}, but in any case de Sitter should be a solution to \eqref{Act} if \(\Lambda\neq0\). If \(\Lambda=0\) then we recover the trivial Minkowski metric written in spherical coordinates.} for which \(A(r)\) is an infinite series, while \(B(r)\) terminates. 
Thus, we see there are solutions valid \( \forall r\in (0,\infty)\), that can be represented by an infintite series. Furthermore, we know the solutions are valid \(\forall r>0\), even if the series expansion of \(A(r)\) or \(B(r)\) at \(r=0\) is not convergent for arbitrary values of \(r\).
This means that the series solutions we find might admit a closed-form that is valid outside the convergence radius of the series solution, making the found solutions even potentially physically relevant outside the convergence region.

More so, we could have a series solution with a finite radius of convergence but such that, a solution for the entire space can be built by matching the solution on the interior region to another asymptotic solution valid for arbitrarily large values of \(r\).
The concept of matching is not new, see for example \cite{Alvarez2020, Stelle2015}. 

This leads to the conclusion that, even if we are not able to obtain the general terms for the series, or even if it is just a valid solution on a reduced region of spacetime, it can be used to build models of static spherically symmetric spacetimes for larger regions of spacetime, provided one can match the series solution to an exterior one.
\dynsection{Necessary conditions for \((s,t)\) families.}\label{sec:3.2}
We will sepparately study the case \((0,0)\) and \((s,t)\neq(0,0)\). 

In the latter case, there are terms in the expansion only proportional to the lowest order coefficients, \(a_s,b_t\neq0\) that are not affected by the higher-order terms in the power series expansions for \(A(r),\; B(r)\).
\\
Then, for the EoMs to be equal to zero, these coefficients need to vanish. This provides a necessary condition for \((s,t)\) to be the consistent solutions to the EoMs.
\\
On dimensional grounds, one can see that these coefficients are proportional to the parameters of the highest curvature operators.
\\
This is the reason why, the quartic action, with two parameters \((\alpha,\beta)\), has more restrictive conditions over \((s,t)\), while the Weyl cubed case, with just a single parameter \(\omega\), has fewer constraints as we show in \cref{Tab:tab1}.

\begin{itemize}
\item If \(\omega\neq0\) the coefficients are, 
    \begin{align}
        &a_s^4 b_t^6 r^{4 s+6 t} \Bigg[6 a_s r^s \Bigg(r^2 \Bigg(\alpha  (t-2) \left(12 s^3-11 s^2 (t-2)-2 s (t-2)^2+(t-2)^3\right)-\nonumber\\
        -&3 \beta  \Big(12 s^3 (t+4)+s^2 \left(-11 t^2+8 t+64\right)-2 s \left(t^3+18 t^2+24 t+16\right)+t^4+4 t^3-\nonumber\\
        -&12 t^2-32 t-80\Big)\Bigg)+(t-2) \omega  \Big(12 s^3+s^2 (62-11 t)-2 s \left(t^2+16 t-52\right)+t^3-\nonumber\\
        -&6 t^2-20 t+56\Big)\Bigg)+(t-2)^2 \omega  (s-t+2)^2 \left(30 s^2+s (t+52)-t^2+4 t+20\right)+\nonumber\\
        +&48 a_s^2 r^{2 s} \left(3 \gamma  r^4 (s-1)-36 \beta  r^2+2 (s+4) \omega \right)+16 a_s^3 r^{3 s} \left(9 \gamma  r^4-6 r^2 (\alpha -3 \beta )-2 \omega \right)\Bigg]\label{c.1.1}.
    \end{align}
    And,
    \begin{align}
        &a_s^3 b_t^6 r^{3 (s+2 t)} \Bigg[6 a_s r^s \Bigg(r^2 \Big(3 \beta  \left(-3 s^2 (t+4)^2+2 s (t-2) (t+4)^2+t^4+4 t^3+36 t^2+64 t+112\right)-\nonumber\\
        -&\alpha  (t-2)^2 \left(-3 s^2+2 s (t-2)+(t-2)^2\right)\Big)-(t-2)^2 \omega  \left(-3 s^2+2 s (t-6)+t^2+4 t-12\right)\Bigg)-\nonumber\\
        -&16 a_s^3 r^{3 s} \left(9 \gamma  r^4-6 r^2 (\alpha -3 \beta )-2 \omega \right)+48 a_s^2 r^{2 s} \left(3 \gamma  r^4 (t+1)-36 \beta  r^2+2 (t-2) \omega \right)+\nonumber\\
        +&(t-2)^3 \omega  (s-t+2)^2 (5 s+t+4)\Bigg].\label{c.2.1}
    \end{align}
    Now, in  \eqref{c.1.1} and \eqref{c.2.1} we can see that, if \(s<0\) the lowest order coefficient will be the term proportional to \(r^{3s}\), this will not recieve additional contributions from the higher order terms in the series \(a_{s+1},b_{t+1},\cdots\). 
   
    Then, since each coefficient must vanish, this would require \(\omega=0\), which is a contradiction. Then, we have our first condition, 
    \begin{equation}
            \bullet \mbox{If } \omega\neq0\Rightarrow s>0.\label{con.1}
    \end{equation}
        Using \eqref{con.1}, we are left with the system,
    \begin{align}
            &a_s^4 b_t^6 r^{4 s+6 t} \Bigg[(t-2)^2 \omega  (s-t+2)^2 \left(30 s^2+s (t+52)-t^2+4 t+20\right)\Bigg]=0,\label{c.1.1.1}\\
            \nonumber\\
            &a_s^3 b_t^6 r^{3 (s+2 t)} \Bigg[(t-2)^3 \omega  (s-t+2)^2 (5 s+t+4)\Bigg]=0.\label{c.2.1.1}
    \end{align}
    Solving this along with \eqref{con.1} yields,
    \begin{equation}
        \bullet \mbox{If } \omega\neq0\Rightarrow \left\lbrace\begin{array}{c} (s,2) \\ (s,2+s) \end{array}\right.\qquad\mbox{for }s\in\IN^+.\label{con.2.2}
    \end{equation}
\item If \(\omega=0\) the coefficients are, 
    \begin{align}
        &6 a_s^5 b_t^6 r^{5 s+6 t+2} \Bigg[-8 a_s^2 r^{2 s} \left(2 \alpha -6 \beta -3 \gamma  r^2\right)-24 a_s r^s \left(12 \beta -\gamma  r^2 (s-1)\right)-\nonumber\\
        -&3 \beta  \Bigg(12 s^3 (t+4)+s^2 \left(-11 t^2+8 t+64\right)-2 s \left(t^3+18 t^2+24 t+16\right)+t^4+4 t^3-\nonumber\\
        -&12 t^2-32 t-80\Bigg)+\alpha  (t-2) \left(12 s^3-11 s^2 (t-2)-2 s (t-2)^2+(t-2)^3\right)\Bigg],\label{c.1.2}
    \end{align}
    \begin{align}
        &6 a_s^4 b_t^6 r^{4 s+6 t+2} \Bigg[-8 a_s^2 r^{2 s} \left(2 \alpha -6 \beta -3 \gamma  r^2\right)+24 a_s r^s \left(12 \beta -\gamma  r^2 (t+1)\right)+\nonumber\\
        +&3 \beta  \left(3 s^2 (t+4)^2-2 s (t-2) (t+4)^2-t^4-4 t^3-36 t^2-64 t-112\right)+\nonumber\\
        +&\alpha  (t-2)^2 \left(-3 s^2+2 s (t-2)+(t-2)^2\right)\Bigg].\label{c.2.2}
    \end{align}
    in this case we have that, \(s\in\mathbb{Z}^+\) and we have the following solutions\footnote{There is one aditional solution already pointed out by Stelle \cite{Stelle2015}, corresponding to \\
    \(s=\frac{t-2}{3}\in\mathbb{Z}^+,\;\alpha=\frac{3 \beta  \left(t^2+2 t+10\right)^2}{(t-2)^4},\;t\geq 3 \).\\
    We do not consider this particular solution in the following, as we do not include in our treatment a case where \(\alpha, \beta\) are functions of \(s,t\).},
    \begin{itemize}
    \item If \(\alpha,\,\beta\neq 0\)
        \begin{equation}
        \bullet \mbox{If }\, \alpha,\,\beta\neq 0 \;\& \;\omega=0\Rightarrow \left\lbrace\begin{array}{c} (1,-1) \\ (2,2) \end{array}\right.,
        \end{equation}
    \item \(\alpha=0\)
        \begin{equation}
        \bullet \mbox{If }\, \beta\neq 0 \;\& \;\omega=\alpha=0\Rightarrow \left\lbrace\begin{array}{c} s=\frac{t^2+2 t+4}{t+4}\in\mathbb{Z}^+\end{array}\right.,
        \end{equation}
    \item \(\beta=0\)
        \begin{equation}
        \bullet \mbox{If }\, \alpha\neq 0 \;\& \;\omega=\beta=0\Rightarrow \left\lbrace\begin{array}{c} s=\frac{2-t}{3}\in\mathbb{Z}^+ \\ s=t-2\\ t=2,\,s \end{array}\right..
        \end{equation}
    \end{itemize}
\end{itemize}
All this information is resumed in \cref{Tab:tab1}.
\begin{table}[h!]
    \begin{center}
        \begin{tabular}{ccccccl}
            \hline
                                                                      & \textbf{$\omega\neq 0$}        & \multicolumn{5}{c}{\textbf{$\omega=0$}}                                                                                                                                                                                          \\ \cline{2-7} 
            \multicolumn{1}{c|}{\multirow{4}{*}{\textbf{(s,t) Families}}} & \multicolumn{1}{l|}{}          & \multicolumn{1}{c|}{\textbf{$\beta\neq0,\;\alpha=0$}}             & \multicolumn{1}{c|}{\textbf{$\beta=-\frac{2\alpha}{3}$}} & \multicolumn{1}{c|}{\textbf{$\beta,\alpha\neq0$}} & \multicolumn{2}{c}{{ $\beta=0,\alpha\neq0$}}    \\ \cline{3-7} 
            \multicolumn{1}{c|}{}                                         & \multicolumn{1}{c|}{$(s,2)$}   & \multicolumn{1}{c|}{\multirow{3}{*}{$(\frac{t^2+2 t+4}{t+4},t)$}} & \multicolumn{1}{c|}{$(1,-1)$}                            & \multicolumn{1}{c|}{$(1,-1)$}                     & \multicolumn{2}{c}{{$(s,2-3s)$}} \\
            \multicolumn{1}{c|}{}                                         & \multicolumn{1}{c|}{$(s,2+s)$} & \multicolumn{1}{c|}{}                                             & \multicolumn{1}{c|}{$(2,2)$}                             & \multicolumn{1}{c|}{$(2,2)$}               & \multicolumn{2}{c}{$(s,2)$}                          \\
            \multicolumn{1}{c|}{}                                         & \multicolumn{1}{c|}{}  & \multicolumn{1}{c|}{}                                             & \multicolumn{1}{c|}{}                            & \multicolumn{1}{c|}{}                                     & \multicolumn{2}{c}{$(s,2+s)$}                         
        \end{tabular}
        \caption{Summary of the non-zero \((s,t)\) families for the different values of the parameters \(\alpha, \,\beta, \,\omega\). If \(\alpha=\omega=\beta=0\) the action corresponds to Einstein-Hilbert with a Cosmological Constant, for which solutions are unique by virtue of the Birkhoff theorem \cite{Schleich2010}.}
        \label{Tab:tab1}
    \end{center}
\end{table}

\levelup{}
\dynsection{Solutions to quadratic and cubic actions.}\label{sec:4}
In this section we explore power series solutions for different combinations of the parameters \(\{\alpha,\,\beta,\,\gamma,\,\omega\}\) in action \eqref{Act}.
In \cref{sec:4.2} we explore the quadratic actions, \(\omega=0\).
Here we found the solutions already presented in \cite{Stelle1978, Stelle2015, Alvarez2020} for the general \(\alpha,\beta\neq0\) case. 
Additionally, we explore in detail some restricted cases of the parameters \(\alpha,\,\beta\), already commented without much depth in \cite{Stelle2015}.

Then, in section \ref{sec:4.3} the presence of a Weyl cubic term is studied. 
To our knowledge, this is a previously unexplored case, with the physical motivation that the on-shell divergent part of pure gravity to two loops is precisely the Weyl cubic tensor as shown in \cite{Goroff1985}. 

As we will see, the (2,2)  solution first found for quadratic gravity, is still a solution for all the cases where \(\omega\neq0\). 

To shorten the expressions, we will leave \(\omega \kappa^2\) of \cref{sec:1} to correspond to \(\omega\) in this section.

This section concludes with a summary of all the obtained solutions in \cref{sec:4.4} and a discussion regarding the presence of singularities at \(r=0\) in \cref{sec:4.5}.

In what follows, we check the EoMs for the families contained in \cref{Tab:tab1}.
\leveldown{}

\dynsection{Solutions for the Quadratic Action, $\omega=0$.}\label{sec:4.2}
In this section, we consider the solutions for an action whose higher curvature operators are quadratic. 
\begin{equation}
    S=\int d^4 x \sqrt{|g|} \, \Big\{-\Lambda-\g \, R-2 \a \, R_{\mu\nu}^2+\left(\b+{2\over 3} \a \right)\,R^2\Big\},\label{Act.Q1}
\end{equation}
We will start, in \ref{sec:4.2.1} with the most complete case, where \(\alpha,\,\beta\neq0\). 
After this, we consider in sections \ref{sec:4.2.2}, \ref{sec:4.2.3}, \ref{sec:4.2.4}, the particular cases for the parameters \(\alpha,\,\beta\).
\leveldown{}
\dynsection{\(\omega=0,\;\alpha,\,\beta\neq 0\).}\label{sec:4.2.1}
This is the case studied in \cite{Stelle2015, Alvarez2020}, obtaining three solution families, 
\begin{equation}
    (0,0),\hspace{1cm}(1,-1),\hspace{1cm}(2,2). \nonumber
\end{equation}
These are reproduced here for completeness.

\begin{enumerate}
\item
The $(0,0)$ family 
\begin{align} &A(r)=1+a_2 r^2+r^4\Bigg\{\frac{\Lambda(3\b-4\a)}{1080\a\b}+ \dfrac{1}{180 \a \b} \Bigg[a_2[\g(2\a+3\b)+ 18\b a_2(10\a+3\b)]+ \nonumber\\
&+ b_2[2\g(-\a+3\b)-18\b b_2(2\a+3\b)]-36\a\b a_2 b_2 \Bigg]\Bigg\}+\mathcal{O}(r^{5}),\label{54}\\&\nonumber\\
&\frac{B(r)}{b_0}=1+b_2 r^2+r^4\Bigg\{\frac{\Lambda(2\a+3\b)}{2160\a\b}+\dfrac{1}{360 \a \b} \Bigg[ a_2[\g(-\a+3\b)+54\b^2 a_2] + \nonumber\\
& +b_2[\g(\a+6\b) +54\b b_2(2\a-\b)]+108\a\b a_2 b_2\Bigg]\Bigg\}+\mathcal{O}(r^{5}).\label{55}
\end{align}
These solutions depend on two parameters $( a_2, b_2)$.
\item
The second family of solutions is the $(1,-1)$ for which the functions read
\begin{align} 
&A(r)=a_1 r-a_1^2 r^2+a_1^3 r^3+a_4 r^4-r^5\frac{a_1}{16}\left(3a_1 b_2+19a_1^4+35a_4\right)+\nonumber\\
&+\frac{a_1^2}{40}\left(21a_1 b_2+101a_1^4+141a_4\right)r^6 + \mathcal{O}(r^{7}), \label{56}  \\
\nonumber\\
&\frac{B(r)}{b_{-1}}=\frac{1}{r}+a_1+b_2 r^2+ r^3\frac{1}{16}\left(a_1 b_2+a_1^4+a_4\right)-r^4\frac{3a_1}{40}\left(a_1 b_2+a_1^4+a_4\right)+\nonumber\\
&+r^5\Bigg\{\frac{(2 \a + 3 \b) \Lambda
    a_1^2}{3888\a\b}+\frac{1}{25920\a \b
    a_1^2} \Bigg[-30\g (\a - 
3 \b)  a_1^6
+ 
81\b(19 \a + 15 \b) 
a_1^8+\nonumber\\
&+ 
162 \b(7 \a + 15 \b) 
a_1^4 a_4- 
405\b (\a - 3 \b) 
a_4^2 + 
10\g (5 \a + 21 \b)
a_1^3 b_2 + \nonumber\\
&+
54\b (161 \a - 15\b) 
a_1^5 b_2 + 
+270 \b(25\a - 3 \b)a_1 a_4 b_2 -\nonumber\\
&- 15 
a_1^2 \left[2 \g(\a- 
3 \b) a_4 + 
9 \b (-53 \a+ 15\b) 
b_2^2\right]\Bigg]\Bigg\}+\mathcal{O}(r^{6}).\label{57}  
\end{align}
These solutions depend on three parameters $( b_2, a_1, a_4)$. The expansion of the Schwarzschild-de Sitter metric around the origin belongs to this family of solutions. 
One can check, substituting in \cref{56,57} the values of $( b_2, a_1, a_4)$ corresponding to the series expansion of the Schwarzschild-de Sitter solution that the coefficients match those of the series expansions in \cref{sec:3.3}.

\item 
Finally, we have the $(2,2)$ family, given by,
\begin{align} &A(r)=a_2 r^2+a_2b_3 r^3-r^4\frac{a_2}{6}\left(2a_2+b_3^2-8b_4\right)+a_5 r^5+\nonumber\\
&+\frac{r^6}{1296\a\b}\Bigg\{-12\a^2a_2^3-2a_2^2\left[b_3^2\left(\a^2-603\a\b-252\b^2\right)+27\a\left(20\b b_4+\g\right)\right]+\nonumber\\
&+a_2\Big[b_3^4\left(-16\a^2+1413\a\b-72\b^2\right)+2b_4b_3^2\left(19\a^2-2223\a\b+180\b^2\right)-\nonumber\\
&-36b_5 b_3\left(\a^2+45\b^2\right)+12\a b_4^2\left(\a+162\b\right)\Big]+324a_5\b b_3\left(7\a+3\b\right)\Bigg\}+\mathcal{O}(r^{7}),\\
 \nonumber \\
&\frac{B(r)}{b_2}=r^2+b_3 r^3+b_4 r^4+b_5 r^5+\frac{r^6}{216\a a_2}\Bigg\{-12\a a_2^3+a_2^2\left[14b_3^2(2\a+3\b)-24\a b_4\right]+\nonumber\\
&+a_2\Big[2b_3^4(67\a-3\b)+2b_4 b_3^2(15\b-227\a)+45b_5 b_3(7\a-3\b)+180\a b_4^2\Big]+\nonumber\\
& +27a_5b_3(\a+3\b)\Bigg\}+ \mathcal{O}(r^{7}).
\end{align}
These solution depend on five parameters $(b_3, b_4, b_5, a_2, a_5)$. 
The Cosmological Constant appears on the coefficients at  order \(\mathcal{O}(r^{10})\). 
\end{enumerate}
\dynsection{\(\omega=0,\;\alpha=0,\,\beta\neq 0\).}\label{sec:4.2.2}
The action, formed by powers of $R$,
    \begin{equation}
    S=\int d^4 x \sqrt{|g|} \, \Big\{-\Lambda-\g \, R+\b\,R^2\Big\},
    \end{equation}
    has not, to our knowledge, ever been subject to a power series analysis like the one we are considering\footnote{In \cite{Stelle2015}, there is a footnote referring to the existence of solutions for special cases of \(\alpha, \;\beta\) but they are not studied in detail.}.
    In \cite{Kehagias2015,Myung2013,Whitt1985,Nelson2010}, Black Hole solutions to pure quadratic gravity are studied and in some cases power series solutions for a different metric gauge are considered.
    
    To find the potential families, we solve the equation
    \begin{equation} 
    s=\frac{4+2t+t^2}{4+t}\qquad\mbox{with }\;\; s,t\in \mathbb{Z}\;\&\;s>0.\label{4.2.2.1}
    \end{equation}
    Obtaining, 
    \begin{align}
        &(s= 1,t= -1),(s= 1,t= 0),(s= 2,t= -2),\nonumber\\&(s= 2,t= 2),(s= 7,t= -3),(s= 7,t= 8).\label{4.2.2.2}
    \end{align}
    We now present further details concerning each of the solutions in \eqref{4.2.2.2}.
    \begin{itemize}
        \item  \((1,-1)\)
        
        In this case, \(\alpha=0\) means that for two orders the EoMs are not linearly independent, and the number of parameters is increased to 6.
        Furthermore, the solution was checked to recover Schwarzwschild-de Sitter coefficients.
        \begin{align}
            &A(r)=a_1 r-a_1^2r^2+a_1^3 r^3-\frac{1}{27 \beta }\Bigg\{9 a_1 \beta  b_2-27 a_1^4 \beta \mp\nonumber\\
            &\mp a_1^2 \gamma\sqrt{a_1^2 \left(-144 a_1 \beta  b_2 \gamma +a_1^2 \left(\gamma ^2-12 \beta  \Lambda \right)+1296 \beta ^2 b_2^2\right)} \Bigg\}r^4+a_5r^5\nonumber\\
            &+a_6r^6+a_7r^7+\mathcal{O}(r^8),\\
            &\frac{B(r)}{b_{-1}}=\frac{1}{r}+a_1+b_2r^2+\frac{4 a_1^5 a_1^2 b_2+7 a_4 a_1 +3 a_5}{7 a_1}r^3+\nonumber\\
            &+\frac{4 a_1^6 -6 a_2^3 b_2+35 a_4 a_1^2 +52 a_5 a_1+21 a_6 }{63 a_1}r^4+\frac{1}{74844 a_1 \beta ^2}\Bigg\{112 a_1^5 \beta  \gamma -\nonumber\\
            &-432 a_1^7 \beta ^2+4176 a_1^4 \beta ^2 b_2+108 a_1^2 \beta  \left(277 a_5 \beta +35 b_2 \gamma \right)-7 a_1^3\Bigg[7 \gamma ^2-\nonumber\\
            &-60 \beta  \Lambda -16 \beta  \sqrt{a_1^2 \left(-144 a_1 \beta  b_2 \gamma +a_1^2 \left(\gamma ^2-12 \beta  \Lambda \right)+1296 \beta ^2 b_2^2\right)}+\nonumber\\
            &+7 a_1 \left(7128 a_5 \beta ^2-7 \gamma  \sqrt{a_1^2 \left(-144 a_1 \beta  b_2 \gamma +a_1^2 \left(\gamma ^2-12 \beta  \Lambda \right)+1296 \beta ^2 b_2^2\right)}-324 b_2^2 \beta ^2\right)+\nonumber\\
            &+252 \beta  \left(81 a_7 \beta -2 b_2 \sqrt{a_1^2 \left(-144 a_1 \beta  b_2 \gamma +a_1^2 \left(\gamma ^2-12 \beta  \Lambda \right)+1296 \beta ^2 b_2^2\right)}\right)\Bigg\}r^5+\mathcal{O}(r^6).
        \end{align}
        We can see that, the two functions above have five free parameters \((a_1,a_5,a_6,a_7,b_2)\).
        
        \item  \((2,-2)\), a 5 parameter\footnote{To sixth order, it increases to six for seventh order.} \((a_2,a_3,a_5,a_6,b_0)\) family.
         
        The first few coefficients in the series for \(A(r)\) and \(B(r)\) are,
        \begin{align}
            &A(r)=a_2 r^2+a_3 r^3+\frac{a_3^2-a_2^3}{a_2}r^4+a_5r^5+a_6r^6+\mathcal{O}(r^7),\label{2m21}\\
            &\frac{B(r)}{b_{-2}}=\frac{1}{r^2}-\frac{a_3 }{a_2}\frac{1}{r}+ b_0+\frac{2 a_3 a_2^3 +a_5 a_2^2-a_3^3 }{a_2^3}r+\nonumber\\
            &+\frac{1}{16 a_2^4 \beta }\Bigg\{a_2^5  \left( \gamma -4 \beta  b_0\right)-8 a_2^6 \beta  +4 a_2^4 \beta  b_0^2+\nonumber\\
            &+\left(8 a_6-11 a_3^2\right) a_2^3 \beta  -a_3 a_2^2 \beta  \left(18 a_5 +a_3 b_0\right)+10 a_3^4 \beta  \Bigg\}r^2+\mathcal{O}(r^3)\label{2m22}
        \end{align}
        where we have the five free parameters, \((a_2,a_3,a_5,a_6,b_0)\).
        This family of solutions is to low order, where only terms proportional to \(\beta\) determine the EoMs, consistent with the solution found in \cite{Kehagias2015}, for pure quadratic gravity, 
        \begin{equation}
            \d s^2=\left(1-\frac{M}{r}\right)^2\d t^2-\left(1-\frac{M}{r}\right)^{-2}\d r^2-r^2\d\Omega_2\label{solpurquad}.
        \end{equation}
        But, when we consider the \(b_3\) coefficient this solution is no longer an expansion of \eqref{solpurquad} as this coefficient now depends on the part proportional to the Ricci scalar of the action.
        \item \((0,0)\)
        \begin{align}
            &A(r)=1+a_2 r^2+\frac{ \left(6 b_2 \left(\gamma -4 \alpha  a_2\right)+ \left(96 \alpha  a_2^2+\Lambda \right)\right)}{120 \alpha }r^4+\nonumber\\
            &+\frac{1}{6720 \alpha ^2 }\Bigg\{ 3 a_2  \left(128 \alpha ^2 b_2^2+ \left(40 \alpha  \Lambda +\gamma ^2\right)+236 \alpha  b_2  \gamma \right)+\nonumber\\
            &+3936 \alpha ^2 a_2^3 -36 \alpha  a_2^2  \left(72 \alpha  b_2+\gamma \right)+b_2 \Bigg(192 \alpha ^2 b_2^2+ \left(3 \gamma ^2-16 \alpha  \Lambda \right)-\nonumber\\
            &-72 \alpha  b_2  \gamma \Bigg)\Bigg\}r^6+\mathcal{O}(r^8), \\
            &\frac{B(r)}{b_0}=1+b_2r^2+\frac{\left(a_2 +b_2\right) \left( \left(\gamma -8 \alpha  a_2\right)+32 \alpha  b_2\right)}{80 \alpha  }r^4+\nonumber\\
            &+\frac{1}{6720 \alpha ^2 }\Bigg\{ 3 a_2  \left(128 \alpha ^2 b_2^2+ \left(40 \alpha  \Lambda +\gamma ^2\right)+236 \alpha  b_2  \gamma \right)+\nonumber\\
            &+3936 \alpha ^2 a_2^3 -36 \alpha  a_2^2  \left(72 \alpha  b_2+ \gamma \right)+b_2 \Bigg(192 \alpha ^2 b_2^2+ \left(3 \gamma ^2-16 \alpha  \Lambda \right)-\nonumber\\
            &-72 \alpha  b_2 \gamma \Bigg)\Bigg\}r^6+\mathcal{O}(r^8), 
        \end{align}
        with two free parameters \((a_2,b_2)\).
        \item  \((1,0)\), a 4 parameter \((a_1,a_2,a_4,b_2)\) family.
        \begin{align}
            &A(r)=a_1 r+a_2 r^2+\frac{45 a_1^2 b_2-32 a_1^4 -88 a_2 a_1^2 -20 a_2^2 }{36 a_1 } r^3+a_4r^4+\mathcal{O}(r^5),\\
            &\frac{B(r)}{b_{0}}=1+\frac{4 \left(a_1^2 +a_2\right)}{3 a_1}r+b_2r^2+ \frac{1}{189a_1^3}\Bigg\{9 a_1^4  \gamma/\beta-160 a_1^6-408 a_2 a_1^4+\nonumber\\
            & +279 a_1^4 b_2-192 a_2^2 a_1^2 +108 a_4 a_1^2 +117 a_2 a_1^2   b_2-52 a_2^3 \Bigg\}r^3+\mathcal{O}(r^4)
        \end{align}
        \item  \((2,2)\), a 5 parameter \((a_2,a_3,a_5,a_6,b_4)\) family.
        This is the  \((2,2)\) family that will be common to all cases, the first few coefficients read, 
        \begin{align} &A(r)=a_2 r^2+a_3 r^3+\frac{8 a_2^2 b_4-2 a_2^3-a_3^2}{6 a_2 }r^4+a_5 r^5+a_6r^6+\mathcal{O}(r^{7})\\
            &\frac{B(r)}{b_2}=r^2+\frac{a_3}{a_2} r^3+b_4 r^4+\frac{14 a_3 a_2^3 +27 a_5 a_2^2 +10 a_3 a_2^2 b_4-2 a_3^3 }{45 a_2^3} r^5+\nonumber\\
            &+\Bigg[\frac{88 a_2  b_4-16 a_2^2+24  b_4^2+34 a_3^2 a_2^{-1} +144 a_6 a_2^{-1}+36 a_3 a_5 a_2^{-2} }{288 }- \nonumber\\
            &-\frac{18 a_3^2 a_2^{-2}  b_4+3 a_3^4a_2^{-4} +6 a_2 \gamma/\beta}{288}\Bigg]r^6 +\mathcal{O}(r^{7}).
        \end{align}
    \end{itemize}
    From the above cases, we see that the higher the values of \(s,t\) the more terms are only determined by the \(\beta R^2\) term.
    This results on the last two families not showing \(\gamma \) dependence at least in the first eight coefficients of the EoMs. This fact impeded us from checking the consistency of these last two solutions, i.e., we did not find coefficients depending on \(\gamma\) thus we can not assert the consistency or inconsistency of this two last families.
    With this situation in mind, we give, for completeness the first few coefficients for each solution.
    \begin{itemize}
        \item  \((7,-3)\), a 5 parameter family.
        \begin{align}
            &A(r)=a_7r^7+a_8r^8+a_9r^9+a_{10}r^{10}+\mathcal{O}(r^{11}),\\
            &\frac{B(r)}{b_{-3}}=\frac{1}{r^3}-\frac{a_8}{9a_7}\frac{1}{r^2}-\frac{91 a_8^3-240 a_7 a_9 a_8+189 a_7^2 a_{10}}{945 a_7^3}\frac{1}{r}+b_0+\mathcal{O}(r).
        \end{align}
        \item  \((7,8)\), a 5 parameter family.
        \begin{align}
            &A(r)=a_7r^7+a_8r^8+\frac{3 \left(2873 a_7^3 b_{10}-2028 a_{10} a_7^2 +16 a_8^3 \right)}{364 a_7 a_8 }r^9+a_{10}r^{10}+\mathcal{O}(r^{11}),\\
            &\frac{B(r)}{b_{8}}=r^8+\frac{12 a_8 }{13 a_7}r^9+\frac{4 \left(6 a_8^2 +169 a_7a_9 \right)}{845 a_7^2}r^9+b_{10}r^{10}+b_{11}r^{11}+\mathcal{O}(r^{12}).
        \end{align}
    \end{itemize}
    We stress that, to this order, there is no dependence on the parameters \(\gamma, \,\beta\), thus consistency of solutions cannot be claimed.
\dynsection{\(\omega=0,\;\alpha\neq0,\,\beta= 0\).}\label{sec:4.2.3}
This action corresponds to the Einstein-Hilbert action with a Weyl squared term;
        \begin{equation}
        S=\int d^4 x \sqrt{|g|} \, \Big\{-\Lambda-\g \, R- \a \, W_{\mu\nu\rho\sigma}W^{\mu\nu\rho\sigma}\Big\}.
        \end{equation}
        In this case, we have,
        \begin{itemize}
        \item  \((n,2)\). For \(n=2\) we have the \((2,2)\), see below. 
        For values \(n\in\{0,1,3,4\}\)  the EoMs imply \(\gamma=0\). The only difference with increasing the value of \(n\) is that the inconsistency appears at increasing order in the EoMs.
        \item  \((n,2-3n)\), was checked for fewer values as \(t=2-3n\) gets large and negative quickly, making the computational task difficult. 
        For \(n\in\{0,2\}\) solutions are not consistent but for \(n=1\) we have the \((1,-1)\) family. 
        While, if \(n\neq1\) the EoMs imply \(\gamma=0\).
        As \(n\) gets larger, the order in which the EoMs show the inconsistency increases.
        \item  \((n,2+n)\). This family was tested for values \(n\in\{0,1,2,3,4\}\). In all cases it required \(\gamma\) to vanish. 
        
        Hence the only solutions found for this action are.
        \item \((0,0)\)
        
        With the corresponding expansions,
        \begin{align}
            &A(r)=1+r^2 \left(b_2+\frac{\Lambda }{3 \gamma }\right)+\frac{ \Lambda  \left(40 \alpha  \Lambda +3 \gamma ^2\right)+18 b_2 \gamma  \left(12 \alpha  \Lambda +\gamma ^2\right)+216 \alpha  b_2^2 \gamma ^2}{360 \alpha  \gamma ^2}r^4+\nonumber\\
            &+a_6 r^6+\mathcal{O}(r^7).
        \end{align}
        And,
        \begin{align}
            &\frac{B(r)}{b_0}=1+b_2 r^2+\frac{ \left(24 \alpha  b_2+ \gamma \right) \left(6 b_2 \gamma + \Lambda \right)}{240 \alpha   \gamma }r^4+\frac{1}{{362880 \alpha ^2  \gamma ^3}}\Bigg\{120960 \alpha ^2 b_2^3 \gamma ^3 +\nonumber\\
            &216 \alpha  b_2^2  \gamma ^2 \Bigg(420 \alpha  \Lambda +97 \gamma ^2\Bigg)+  \Bigg[\Lambda  \left(2240 \alpha ^2 \Lambda ^2+732 \alpha  \gamma ^2 \Lambda +27 \gamma ^4\right)-\nonumber\\
            &-60480 \alpha ^2 a_6 \gamma ^3\Bigg]+6 b_2  \gamma  \Bigg(3920 \alpha ^2 \Lambda ^2+1314 \alpha  \gamma ^2 \Lambda +27 \gamma ^4\Bigg)\Bigg\}r^6+\mathcal{O}(r^8),
        \end{align}
        with the two free parameters, \((a_6,b_2)\).
        \item \((2,2)\)
        \begin{align}
            &A(r)=a_2 r^2+a_2 b_3 r^3r+a_4 r^4+a_5 r^5+a_6 r^6+\mathcal{O}(r^7),\\
            &\frac{B(r)}{b_2}=r^2+b_3r^3+b_4 r^4+\frac{1}{48 \alpha  a_2^2 b_3}\Bigg\{-16 \alpha  a_2^4-7 \alpha  a_3^2 b_3^2-72 a_2^3 \gamma+\nonumber\\
            &+\alpha  a_2^2 \left(-7 b_3^4+24 b_4 b_3^2+16 b_4^2\right)+2 \alpha  a_2 b_3 \left(4 a_4 b_3+a_3 \left(8 b_4-3 b_3^2\right)\right)\Bigg\}r^5+\nonumber\\
            &+\frac{1}{96 \alpha  a_2}\Bigg\{ \alpha  a_2 \left(-5 b_3^4-30 b_4 b_3^2+84 b_4^2\right)+2 \alpha  \left(6 a_5 b_3+a_2 \left(16 b_4-7 b_3^2\right)\right)-\nonumber\\
            &-52 \alpha  a_2^3-210 a_2^2 \gamma\Bigg\}r^6-\frac{1}{960 \alpha  a_2^2 b_3}\Bigg\{12 \alpha  a_2^4 \left(21 b_3^2+20 b_4\right)+\nonumber\\
            &+24 \alpha  a_2^2 b_3^2+2 a_2^3 \left(80 \alpha  a_2+351 b_3^2 \gamma +540 b_4 \gamma \right)+\nonumber\\
            &-2\alpha a_2\left(48 a_4 b_3^2-6 a_5 \left(9 b_3^3-20 b_3 b_4\right)+a_2 \left(-37 b_3^4+60 b_4 b_3^2+80 b_4^2\right)\right)+\nonumber\\
            &+5 a_2^2 \left(144 a_2 \gamma -\alpha  \left(21 b_3^6-102 b_4 b_3^4+92 b_4^2 b_3^2+48 b_4^3\right)\right)\Bigg\}r^7+\mathcal{O}(r^8),
        \end{align}
        with six parameters \((a_2,a_4,a_5,a_6,b_3,b_4)\).
        \item \((1,-1)\)        
        Again, the fact that we have set one of the two parameters to zero increases the number of free coefficients in the solutions to 5 free parameters \((a_1,a_5,a_6,a_7,b_2)\)
        \begin{align}
            &A(r)=a_1 r-a_1^2 r^2+a_1^3 r^3+ \frac{ \left(15 a_1 b_2 \gamma -9 a_1^4 \gamma +4 a_1^2 \Lambda \right)}{9 \gamma }r^4+a_5 r^5+a_6 r^6+\nonumber\\
            &+a_7 r^7+\mathcal{O}(r^8),
        \end{align}
        and, 
        \begin{align}
            &\frac{B(r)}{b_{-1}}=\frac{1}{r}+a_1+b_2r^2-\frac{8 a_1^3 b_0 \Lambda \gamma^{-1}-9 a_1^5 b_0+30 a_1^2 b_2+9 a_5 b_0}{27 a_1}r^3-\nonumber\\
            &-\frac{-225 a_1^6 b_0 \gamma +417 a_1^3 b_2 \gamma +468 a_5 a_1 b_0 \gamma +243 a_6 b_0 \gamma +104 a_1^4 b_0 \Lambda }{891 a_1 \gamma }r^4+\nonumber\\
            &+\frac{1}{23166 \alpha  a_1 \gamma ^2}\Bigg\{ \Bigg(33 a_1^3 \Lambda  \left(28 \alpha  \Lambda +3 \gamma ^2\right)-72 \alpha  a_1^7 \gamma ^2-5346 \alpha  a_7 \gamma ^2+\nonumber\\
            &+160 \alpha  a_1^5 \gamma  \Lambda +18 a_1^2 \gamma  \Bigg(11 b_2 \left(44 \alpha  \Lambda +3 \gamma ^2\right)-257 \alpha  a_5 \gamma \Bigg)+\nonumber\\
            &+1464 \alpha  a_1^4 b_2 \gamma ^2-54 \alpha  a_1 \gamma ^2 \left(186 a_6-451 b_2^2\right)\Bigg)\Bigg\}r^5+\mathcal{O}(r^6).
        \end{align}
        \end{itemize}
\dynsection{\(\omega=0,\;\beta+\frac{2}{3}\alpha=0\).}\label{sec:4.2.4}
We have now the following action,
        \begin{equation}
        S=\int d^4 x \sqrt{|g|} \, \Big\{-\Lambda-\g \, R-2 \a \, R_{\mu\nu}^2\Big\},
        \end{equation}
        the \((s,t)\) values satisfying the necessary conditions are; 
        \begin{equation} s=-t=1,\hspace{1cm}s=t=2.\end{equation}
        Thus, we have the solutions,
        \begin{itemize}
            \item \((0,0)\), a 2 parameter \((a_2,b_2)\) family.
            \begin{align}
                &A(r)=1+a_2 r^2+\frac{ \left(6 b_2 \left(\gamma -4 \alpha  a_2\right)+ \left(96 \alpha  a_2^2+\Lambda \right)\right)}{120 \alpha}r^4+\frac{1}{6720 \alpha ^2 }\Bigg\{3 a_2  \Bigg[128 \alpha ^2 b_2^2+ \nonumber\\
                &\left(40 \alpha  \Lambda +\gamma ^2\right)+236 \alpha  b_2  \gamma \Bigg]+3936 \alpha ^2 a_2^3-36 \alpha  a_2^2  \left(72 \alpha  b_2+ \gamma \right)+\nonumber\\
                &+b_2 \left(192 \alpha ^2 b_2^2+ \left(3 \gamma ^2-16 \alpha  \Lambda \right)-72 \alpha  b_2  \gamma \right)\Bigg\}r^6 +\mathcal{O}(r^7),\\
                &\frac{B(r)}{b_0}=1+b_2 r^2-\frac{\left(a_2 +b_2\right) \left( \left(-\left(\gamma -8 \alpha  a_2\right)\right)-32 \alpha  b_2\right)}{80 \alpha  }r^4 +\nonumber\\
                &+\frac{1}{40320 \alpha ^2}\Bigg\{-2496 \alpha ^2 a_2^3+3840 \alpha ^2 b_2^3+b_2 \left(56 \alpha  \Lambda +9 \gamma ^2\right)+768 \alpha  b_2^2 \gamma +\gamma  \Lambda+\nonumber\\
                &+a_2 \left(-16 \alpha  \Lambda +8256 \alpha ^2 b_2^2+648 \alpha  b_2 \gamma +3 \gamma ^2\right)++24 \alpha  a_2^2 \left(80 \alpha  b_2+13 \gamma \right)\Bigg\}r^6+\mathcal{O}(r^8).
            \end{align}

            \item \((1,-1)\), a 3 \((a_1,a_4,b_2)\) parameter family. 
            \begin{align}
                &A(r)=a_1 r-a_1^2 r^2+a_1^3 r^3+a_4 r^4+\frac{1}{16} r^5 \left(-3 a_1^2 b_2-19 a_1^5-35 a_4 a_1\right)+\nonumber\\
                &+\frac{1}{40} a_1^2 r^6 \left(21 a_1 b_2+101 a_1^4+141 a_4\right)+\frac{r^7}{25920 \alpha  a_1}\Bigg\{135 a_1^6 \gamma -81891 \alpha  a_1^8+\nonumber\\
                &+21465 \alpha  a_4^2+2 a_1^4 \left(80 \Lambda -43173 \alpha  a_4\right)+135 a_1^2 \left(a_4 \gamma -25 \alpha  b_2^2\right)-\nonumber\\
                &-33426 \alpha  a_1^5 b_2-7830 \alpha  a_4 a_1 b_2+1095 a_1^3 b_2 \gamma\Bigg\}+\mathcal{O}(r^8),\\
                &\frac{B(r)}{b_{-1}}=\frac{1}{r}+a_1+b_2r^2+\frac{1}{16} r^3 \left(a_1 b_2+a_1^4+a_4\right)-\frac{3}{40} r^4 \left(a_1^2 b_2+a_1^5+a_4 a_1\right)+\nonumber\\
                &+\frac{ \left(a_1 b_2+a_1^4+a_4\right) \left(27 \alpha  a_1^4-45 \alpha  a_4+315 \alpha  a_1 b_2+5 a_1^2 \gamma \right)}{960 \alpha  a_1^2}r^5+\mathcal{O}(r^6).
            \end{align}
            \item \((2,2)\), a 5 parameter \((a_2,a_4,a_6,b_3,b_5)\) family.
            \begin{align}
                &A(r)=a_2r^2+a_2b_3 r^3+a_4 r^4+\frac{1}{1920 \alpha  a_2 b_3}\Bigg\{156 \alpha  a_2^4+104 \alpha  a_4 a_2^2+1536 \alpha  a_6 a_2-\nonumber\\
                &-1284 \alpha  a_4^2+388 \alpha  a_2^3 b_3^2-1093 \alpha  a_2^2 b_3^4-1344 \alpha  a_2^2 b_3 b_5-96 a_2^3 \gamma \nonumber\\
                &+3788 \alpha  a_4 a_2 b_3^2\Bigg\}r^5+a_6r^6+\mathcal{O}(r^7)\\
                &\frac{B(r)}{b_2}=r^2+b_3 r^3+\frac{a_2 b_3^2+2 a_2^2+6 a_4}{8 a_2} r^4 +b_5 r^5 +\frac{1}{15360   a_2^2}\Bigg\{8484  a_4^2-8228  a_2^3 b_3^2+\nonumber\\
                &+6893   a_2^2 b_3^4-636  a_2^4+3416   a_4 a_2^2-1536   a_6 a_2+30144  a_2^2 b_3 b_5-\nonumber\\
                &-26668  a_4 a_2 b_3^2+96 a_2^3 \gamma /\alpha\Bigg\}r^6+\mathcal{O}(r^7)
            \end{align}
        \end{itemize}
\levelup{}
\dynsection{Solutions for the Cubic Action, $\omega\neq0$.}\label{sec:4.3}
As discussed in \cref{sec:3.2}, the potential families are given by, 
\begin{equation}
     (0,0),\hspace{1cm}(s,2),\hspace{1cm}(s,2+s),\hspace{1cm}\mbox{with } \;s\in \mathbb{Z}^+.\label{c.om}
\end{equation}
Notice the Schwarzschild-de Sitter solution \textbf{is not contained in \cref{c.om} above}. 
This feature was pointed out in \cite{Deser2003}, where they used the Weyl cubed operator to build an action that does not admit a Schwarzschild-de Sitter solution.

Now, what remains is to check if the necessary conditions given by \eqref{c.om} correspond to consistent solutions of the EoMs for the different values of \(\alpha\) and \(\beta\).
Let us start from the complete action \eqref{Act}.
\leveldown{}
\dynsection{\(\omega\neq0,\;\alpha,\,\beta\neq 0\).}\label{sec:4.3.1}
Since there are contributions from the Ricci scalar, the Ricci scalar squared and Ricci tensor squared, we would expect that their effect on the potential families of solutions in \eqref{c.om} would be to rule out most of the families, while cases in which one of the parameters could, potentially admit additional solutions.
We will see that the only possible solutions for any combination are indeed the \((0,0)\) and \((2,2)\) families.

In this case we found,
\begin{itemize}
\item
The $(0,0)$ family
\begin{align} 
    &A(r)=1+a_2r^2+\frac{1}{1080\alpha\beta}\Bigg\{1080 \alpha  a_2^2 \beta +12 \alpha  a_2 \gamma +324 a_2^2 \beta ^2-216 \alpha  a_2 \beta  b_2-18 a_2 \beta  \gamma +\nonumber\\
    &-216 \alpha  \beta  b_2^2\Bigg\}r^4+a_6(a_2,b_2,\alpha,\beta,\gamma,\lambda,\omega)r^6+O(r^8),\\
    &\frac{B(r)}{b_0}=1+b_2r^2+\frac{1}{2160\alpha\beta}\Bigg\{324 a_2^2 \beta ^2+6 a_2 \left(-\alpha  \gamma +108 \alpha  \beta  b_2+3 \beta  \gamma \right)+\Lambda  (2 \alpha +3 \beta )+\nonumber\\
    &+6 b_2 \gamma  (\alpha +6 \beta )+324 \beta  b_2^2 (2 \alpha -\beta )\Bigg\}r^4+b_6(a_2,b_2,\alpha,\beta,\gamma,\lambda,\omega)r^6+O(r^8).
\end{align}
With free parameters, \((a_2,b_2)\).
\((a_6,b_6)\) are not explicitly written to save some space, as they are too long even for the standards of this thesis. 
\item
The $(2,2)$ family,
\begin{align}
     &A(r)=a_2r^2+a_2b_3r^3+\frac{1}{2} a_2 \left(2 a_2+b_3^2\right)r^4+a_5r^5+\frac{1}{864 a_2 \beta }\Bigg\{3 a_5 b_3^2 \left(b_3^3-6 b_5\right) \omega-576 a_2^4 \beta-\nonumber\\
     &-4 a_2^3 \left(-2 b_3 \left(b_3 (4 \alpha +105 \beta )+12 b_5 \omega \right)+56 b_3^4 \omega +9 \gamma \right)-4 a_2^2 \Bigg(-b_3^3 \Big(b_3 (\alpha +42 \beta )+\nonumber\\
     &+111 b_5 \omega \Big)+6 b_5 \left(b_3 (\alpha +105 \beta )+3 b_5 \omega \right)-432 \beta  b_6+17 b_3^6 \omega +48 b_6 b_3^2 \omega \Bigg)+\nonumber\\
     &+a_2 b_3 \left(24 a_5 \left(b_3^2 \omega +54 \beta \right)-\left(b_3^3-6 b_5\right) \left(5 b_3^4-33 b_5 b_3+24 b_6\right) \omega \right)\Bigg\}r^6+O(r^7),\\
    &\frac{B(r)}{b_2}=r^2+b_3r^3+\frac{1}{2} \left(2 a_2+b_3^2\right)r^4+b_5r^5+b_6r^6+O(r^7),
\end{align}
with 5 free parameters corresponding to \((a_2,a_5,b_3,b_5,b_6)\).
\item
For the $(n,2)$ family; with \(n\in \mathbb{N}/\{2\}\) the EoMs require \(\gamma\rightarrow0\).
\item  
The $(n,2\,+n)$ family was tested for values \(n=\{0,1,2,3\}\) and the EoMs required the quadratic coefficients \((\alpha,\,\beta)\) to vanish. 
As usual, the order in which the inconsistency appears increases with \(n\).
\end{itemize}
Summing up, for the more general case of action \eqref{Act} we have found that only two families can constitute power series solutions to the equations of motion. 
The \((0,0)\) solution, and the \((2,2)\) solution. Some properties of the two solutions are presented in \cref{sec:4.4}.

\textbf{Note again that the Schwarzschild-de Sitter solution is not a solution to the action } \eqref{Act}.
\dynsection{\(\omega\neq0,\;\alpha=0,\,\beta\neq 0\).}\label{sec:4.3.2}
In this case we have an action form by powers of $R$ plus Weyl cubic
    \begin{equation}
    S=\int d^4 x \sqrt{|g|} \, \Big\{-\Lambda-\g \, R+\b\,R^2+\omega W_{\mu\nu\a\b}W^{\a\b\rho\sigma}W_{\rho\sigma}^{~~\mu\nu}\Big\}.
    \end{equation}
    Again the solutions are the two families, 
        \begin{itemize}
            \item \((0,0)\) 
            \begin{align}
                &A(r)=1+a_2 r^2\pm\frac{\sqrt{\left(-\left(108 a_2^2 \beta +6 a_2 \gamma +\Lambda \right)\right)+108 \beta  b_2^2-12 b_2 \gamma }}{3 \sqrt{15}  \sqrt{\omega }}r^3+\nonumber\\
                &+a_4r^4+\mathcal{O}(r^5),\\
                &\frac{B(r)}{b_0}=1+b_2 r^2 \pm\frac{2}{3}\frac{\sqrt{\left(-\left(108 a_2^2 \beta +6 a_2 \gamma +\Lambda \right)\right)+108 \beta  b_2^2-12 b_2 \gamma }}{3 \sqrt{15}  \sqrt{\omega }}r^3+\nonumber\\
                &+\frac{ \left(-180 a_2^2 \beta +180 a_4 \beta -3 a_2 \gamma +\Lambda \right)+3 b_2  \left(48 a_2 \beta +\gamma \right)+144 \beta  b_2^2}{360 \beta  }r^4+\mathcal{O}(r^5),  
            \end{align}
            with three free parameters \((a_2,a_4,b_2)\)
            \item \((2,2)\) 
            \begin{align}
                &A(r)=a_2r^2+a_2b_3r^3+\left(\frac{1}{2} a_2 \left(2 a_2+b_3^2\right)\right)r^4+a_5r^5+a_6r^6+\mathcal{O}(r^7),\\
                &\frac{B(r)}{b_2}=r^2+b_3r^3+\left(a_2+\frac{b_3^2}{2}\right)r^4+b_5r^5+\nonumber\\
                &+\frac{1}{24 a_2 \left(8 a_2 \left(9 \beta -b_5^2 \omega \right)-b_5^2 \left(b_5^2-6\right) \omega \right)}\Bigg\{576 a_2^4 \beta+4 a_2^3 \left(-6 b_5^2 (35 \beta +4 \omega )+56 b_5^4 \omega +9 \gamma \right)+\nonumber\\
                &+a_2 \left(-24 a_5 \left(54 \beta  b_5+b_5^3 \omega \right)+864 a_4 \beta +\left(5 b_5^4-63 b_5^2+198\right) b_5^4 \omega \right)-3 a_5 b_5^3 \left(b_5^2-6\right) \omega+\nonumber\\
                &+4 a_2^2 b_5^2 \left(-3 b_5^2 (14 \beta +37 \omega )+17 b_5^4 \omega +18 (35 \beta +\omega )\right) \Bigg\}r^6+\mathcal{O}(r^7), 
            \end{align}
            with five free parameters \((a_2,a_5,a_6,b_3,b_5)\).
        \end{itemize}
\dynsection{\(\omega\neq0,\;\alpha\neq0,\,\beta= 0\).}\label{sec:4.3.3}
The action reads, 
\begin{equation}
S=\int d^4 x \sqrt{|g|} \, \Big\{-\Lambda-\g \, R- \a \, W_{\mu\nu\rho\sigma}W^{\mu\nu\rho\sigma}+\omega W_{\mu\nu\a\b}W^{\a\b\rho\sigma}W_{\rho\sigma}^{~~\mu\nu}\Big\}.
\end{equation}
Again the admitted solutions are
\begin{itemize}
    \item \((0,0)\) 
    \begin{align}
        &A(r)=1+a_2 r^2+\frac{r^4 \left(18 a_2 \left(4 \alpha  \gamma  \Lambda +\gamma ^3\right)+216 \alpha  a_2^2 \gamma ^2-\Lambda  \left(8 \alpha  \Lambda +3 \gamma ^2\right)\right)}{360 \alpha  \gamma ^2}+\nonumber\\
        &+6a_6r^6+\mathcal{O}(r^8),\\
        &\frac{B(r)}{b_0}=1-\frac{ \left(\Lambda -3 a_2 \gamma \right)}{3 \gamma }r^2+\frac{\left(6 a_2 \gamma -\Lambda \right) \left(24 \alpha  a_2 \gamma -8 \alpha  \Lambda +\gamma ^2\right)}{240 \alpha  \gamma ^2}r^4+\nonumber\\
        &+\frac{1}{2419200 \alpha ^3 \gamma ^2}\Bigg\{4 a_2 \left(5600 \alpha ^3 \Lambda ^2-10140 \alpha ^2 \gamma ^2 \Lambda +270 \alpha  \gamma ^4-189 \gamma ^3 \Lambda  \omega \right)+44800 a_6 \alpha ^3 \gamma -\nonumber\\
        &-36 a_2^2 \left(5600 \alpha ^3 \gamma  \Lambda -3880 \alpha ^2 \gamma ^3-63 \gamma ^4 \omega \right)-9 \gamma  \Bigg(\Lambda  \Big(-320 \alpha ^2 \Lambda +20 \alpha  \gamma ^2-\nonumber\\
        &-7 \gamma  \Lambda  \omega \Big)+806400 \alpha ^3 a_2^3 \gamma ^2\Bigg)\Bigg\}r^6+\mathcal{O}(r^8),
    \end{align}
    with two free parameters \((a_2,a_6)\).
    \item \((2,2)\) 
    \begin{align}
        &A(r)=a_2r^2+a_3 r^3+a_4 r^4+\mathcal{O}(r^5),\\\
        &\frac{B(r)}{b_2}=r^2+b_3 r^3+\frac{1}{4} \left(\frac{a_3 b_3}{a_2}+4 a_2+b_3^2\right) r^4 +b_5r^5+\mathcal{O}(r^6),
    \end{align} 
    with five free parameters \((a_2,a_3,a_4,b_3,b_5)\).
\end{itemize}
\dynsection{\(\omega\neq0,\;\beta+\frac{2}{3}\alpha=0\).}\label{sec:4.3.4}
We have an action with Ricci squared plus Weyl cubic
\begin{equation}
S=\int d^4 x \sqrt{|g|} \, \Big\{-\Lambda-\g \, R-2 \a \, R_{\mu\nu}^2+\omega W_{\mu\nu\a\b}W^{\a\b\rho\sigma}W_{\rho\sigma}^{~~\mu\nu}\Big\}.
\end{equation}
In this case, the only possible solutions found correspond to 
\begin{itemize}
    \item \((0,0)\) 
    \begin{align}
        &A(r)=1+a_2 r^2+\frac{6 b_2 \left(\gamma -4 \alpha  a_2\right)+\left(96 \alpha  a_2^2+\Lambda \right)}{120 \alpha  }r^4+\frac{1}{1209600 \alpha ^3 }\Bigg\{7 \Lambda ^2 \omega +36288 \alpha ^2 a_2^4 \omega+\nonumber\\
        &+864 a_2^3 \left(820 \alpha ^3-7 \alpha  \gamma  \omega \right)b_2 +\left(-2880 \alpha ^2 \Lambda +540 \alpha  \gamma ^2+168 \gamma  \Lambda  \omega \right)+36288 \alpha ^2 b_2^4 \omega-\nonumber\\
        & -36 a_2^2 \left(180 \alpha ^2 \gamma +28 \alpha  \Lambda  \omega +48 b_2 \left(270 \alpha ^3+7 \alpha  \gamma  \omega \right)+2016 \alpha ^2 b_2^2 \omega -7 \gamma ^2 \omega \right)-\nonumber\\
        &-144 b_2^2 \left(90 \alpha ^2 \gamma -7 \alpha  \Lambda  \omega -7 \gamma ^2 \omega \right)+12 a_2 \Bigg(1800 \alpha ^2 \Lambda +45 \alpha  \gamma ^2+\nonumber\\
        &+72 b_2^2 \left(80 \alpha ^3+7 \alpha  \gamma  \omega \right)+12 b_2 \gamma  \left(885 \alpha ^2+7 \gamma  \omega \right)+7 \gamma  \Lambda  \omega \Bigg)\Bigg\}r^6+\mathcal{O}(r^8),
    \end{align}
    and,
    \begin{align}
        &\frac{B(r)}{b_0}=1+b_2 r^2 +\frac{\left(a_2 +b_2\right) \left(\left(\gamma -8 \alpha  a_2\right)+32 \alpha  b_2\right)}{80 \alpha  } r^4 +\frac{1}{3628800 \alpha ^3}\Bigg\{90 \alpha\Lambda  \gamma -\nonumber\\
        &-864 a_2^3 \left(260 \alpha ^3+7 \alpha  \gamma  \omega \right)+36288 \alpha ^2 a_2^4 \omega +6 b_2 \left(840 \alpha ^2 \Lambda +135 \alpha  \gamma ^2+28 \gamma  \Lambda  \omega \right)+\nonumber\\
        &+6 a_2 \left(-240 \alpha ^2 \Lambda +45 \alpha  \gamma ^2+144 b_2^2 \left(860 \alpha ^3+7 \alpha  \gamma  \omega \right)+24 b_2 \gamma  \left(405 \alpha ^2+7 \gamma  \omega \right)+14 \gamma  \Lambda  \omega \right)+\nonumber\\
        &+36288 \alpha ^2 b_2^4 \omega+144 b_2^2 \left(480 \alpha ^2 \gamma +7 \alpha  \Lambda  \omega +7 \gamma ^2 \omega \right)+1728 b_2^3 \left(200 \alpha ^3+7 \alpha  \gamma  \omega \right)+\nonumber\\
        &+36 a_2^2 \left(780 \alpha ^2 \gamma -28 \alpha  \Lambda  \omega +48 b_2 \left(100 \alpha ^3-7 \alpha  \gamma  \omega \right)-2016 \alpha ^2 b_2^2 \omega +7 \gamma ^2 \omega \right)+\nonumber\\
        &+ 7 \Lambda^2  \omega\Bigg\}r^6+\mathcal{O}(r^8)   ,
    \end{align}
    with two free parameters \((a_2,b_2)\).
    \item \((2,2)\) 
    \begin{align}
        &A(r)=a_2r^2+a_2b_3 r^3+\frac{1}{2} a_2 \left(2 a_2+b_3^2\right) r^4+a_5r^5+\frac{1}{576 \alpha  a_2}\Bigg\{288 \alpha  b_6+\nonumber\\
        &+4 a_2^2 \left(3 b_3^3 \left(9 \alpha  b_3-37 b_5 \omega \right)+18 b_5 \left(b_5 \omega -23 \alpha  b_3\right)+17 b_3^6 \omega +48 b_6 b_3^2 \omega \right)+\nonumber\\
        &+4 a_2^3 \left(12 b_3 \left(11 \alpha  b_3-2 b_5 \omega \right)+56 b_3^4 \omega +9 \gamma \right)3 a_5 b_3^2 \left(6 b_5-b_3^3\right) \omega -384 \alpha  a_2^4+\nonumber\\
        &+a_2 b_3 \left(24 a_5 \left(36 \alpha -b_3^2 \omega \right)+\left(b_3^3-6 b_5\right) \left(5 b_3^4-33 b_5 b_3+24 b_6\right) \omega \right)\Bigg\}r^6+\mathcal{O}(r^7),\\
        &\frac{B(r)}{b_2}=r^2+b_3 r^3+\frac{1}{2} \left(2 a_2+b_3^2\right) r^4 +b_5 r^5 +b_6r^6+\mathcal{O}(r^7),
    \end{align}
    with five free parameters \((a_2,a_5,b_3,b_5,b_6)\).
\end{itemize}
\dynsection{\(\omega\neq0,\;\alpha=\beta=0\).}\label{sec:4.3.5}
In this case there are no quadratic terms, i.e. $\a=\b=0$. This case would correspond to the Renormalization to two loops of pure gravity;
\begin{equation}
S=\int d^4 x \sqrt{|g|} \, \Big\{-\Lambda-\g \, R+\omega W_{\mu\nu\a\b}W^{\a\b\rho\sigma}W_{\rho\sigma}^{~~\mu\nu}\Big\}.
\end{equation}
The consistent solutions correspond to:

\begin{itemize} 
\item The $(0,0)$ family now splits into several different solutions. 
First the de Sitter solution
\begin{align}
    &A(r)=1+\frac{\Lambda }{6 \gamma }r^2+\frac{\Lambda ^2}{36 \gamma ^2}r^4+\frac{\Lambda ^3}{216 \gamma ^3}r^6+\mathcal{O}(r^8),\\
    &\frac{B(r)}{b_0}=1-\frac{\Lambda }{6 \gamma }r^2+\mathcal{O}(r^8).
\end{align}
Then we have with similar structure,
\begin{align}
    &A(r)=1+\frac{\Lambda }{6 \gamma }r^2+\left[\frac{\Lambda ^2}{36 \gamma ^2}-\frac{\gamma }{42 \omega }\right]r^4+\left[\frac{\Lambda ^3}{216 \gamma ^3}-\frac{251 \Lambda }{25578 \omega }\right]r^6+\mathcal{O}(r^8),\\
    &\frac{B(r)}{b_0}=1-\frac{\Lambda }{6 \gamma }r^2-\frac{ \gamma }{84 \omega }r^4+\frac{9  \Lambda }{11368 \omega }r^6+\mathcal{O}(r^8).
\end{align}
Now we have the pair of solutions
\begin{align}
    &A(r)=1-\frac{\Lambda }{2 \gamma }r^2\pm\frac{2 \sqrt{\Lambda }}{3 \sqrt{5} \sqrt{\omega }}r^3+\left[\frac{\Lambda ^2-\frac{\gamma ^3}{\omega }}{36 \gamma ^2}\right]r^4\mp\left[\frac{\Lambda ^3}{216 \gamma ^3}-\frac{25 \gamma ^3+18776 \Lambda ^2 \omega }{45360 \sqrt{5} \gamma  \sqrt{\Lambda } \omega ^{3/2}}\right]r^5-\nonumber\\
    &-\left[\frac{\Lambda ^3}{24 \gamma ^3}-\frac{5 \gamma ^3}{193536 \Lambda  \omega ^2}+\frac{17693 \Lambda }{181440 \omega }\right]r^6 \mp\frac{13875 \gamma ^6+61571560 \gamma ^3 \Lambda ^2 \omega -29049984 \Lambda ^4 \omega ^2}{1828915200 \sqrt{5} \gamma ^2 \Lambda ^{3/2} \omega ^{5/2}}r^7+\nonumber\\
    &+a_8r^8+\mathcal{O}(r^9),
\end{align}
and,
\begin{align}
    &\frac{B(r)}{b_0}=1-\frac{5 \Lambda }{6 \gamma }r^2\pm\frac{4 \sqrt{\Lambda }}{9 \sqrt{5} \sqrt{\omega }}r^3-\left[\frac{ \left(\gamma ^3-24 \Lambda ^2 \omega \right)}{72 \gamma ^2 \omega }\right]r^4+\nonumber\\
    &\mp\left[\frac{\left(25 \gamma ^3+53972 \Lambda ^2 \omega \right)}{113400 \sqrt{5} \gamma  \sqrt{\Lambda } \omega ^{3/2}}\right]r^5-\frac{ \left(75 \gamma ^6-451088 \gamma ^3 \Lambda ^2 \omega +645120 \Lambda ^4 \omega ^2\right)}{8709120 \gamma ^3 \Lambda  \omega ^2}r^6+\nonumber\\
    &\mp\frac{ \left(13875 \gamma ^6+77806660 \gamma ^3 \Lambda ^2 \omega -1487323584 \Lambda ^4 \omega ^2\right)}{6401203200 \sqrt{5} \gamma ^2 \Lambda ^{3/2} \omega ^{5/2}}r^7-\frac{1}{50164531200 \gamma ^4 \Lambda ^2 \omega ^3}\Bigg\{9525 \gamma ^9+\nonumber\\
    &+6270566400 a_8 \gamma ^4 \Lambda ^2 \omega ^3-13367968 \gamma ^6 \Lambda ^2 \omega +3117059328 \gamma ^3 \Lambda ^4 \omega ^2-\nonumber\\
    &-237081600 \Lambda ^6 \omega ^3\Bigg\}r^8+\mathcal{O}(r^9),
\end{align}
with one parameter \(a_8\).

And lastly the two parameter \((a_3,a_8)\) family,
\begin{align}
    &A(r)=1+\frac{\Lambda -45 a_3^2 \omega }{6 \gamma }r^2+a_3r^3-\frac{108 a_3^2 \Lambda  \omega ^2-1215 a_3^4 \omega ^3+\gamma ^3-\Lambda ^2 \omega }{36 \gamma ^2 \omega }r^4-\nonumber\\
    &-\frac{-5490 a_3^2 \Lambda +156816 a_3^4 \omega +\frac{\gamma ^3}{\omega ^2}}{13608 a_3 \gamma }r^5+\frac{1}{435456 a_3^2 \gamma ^3 \omega ^3}\Bigg\{-\gamma ^6+12 a_3^2 \Lambda  \omega ^2 \left(168 \Lambda ^2 \omega -415 \gamma ^3\right)+\nonumber\\
    &+216 a_3^4 \omega ^3 \left(2471 \gamma ^3-1704 \Lambda ^2 \omega \right)+10847520 a_3^6 \Lambda  \omega ^5-52488000 a_3^8 \omega ^6\Bigg\}r^6+\nonumber\\
    &+\frac{1}{411505920 a_3^3 \gamma ^2 \omega ^4} \Bigg\{216 a_3^4 \omega ^3 \left(217701 \Lambda ^2 \omega -95549 \gamma ^3\right)-37 \gamma ^6-\nonumber\\
    &-12606 a_3^2 \gamma ^3 \Lambda  \omega ^231675458240 a_3^8 \omega ^6-3234310560 a_3^6 \Lambda  \omega ^5\Bigg\}r^7+a_8r^8+\mathcal{O}(r^9),
\end{align}
and
\begin{align}
    &\frac{B(r)}{b_0}=1-\frac{45 a_3^2 \omega +\Lambda }{6 \gamma }r^2+\frac{2 a_3 }{3}r^3-\frac{ \left(-54 a_3^2 \Lambda  \omega ^2-2430 a_3^4 \omega ^3+\gamma ^3\right)}{72 \gamma ^2 \omega }r^4-\nonumber\\
    &-\frac{ \left(1881 a_3^2 \Lambda  \omega ^2+252072 a_3^4 \omega ^3+\gamma ^3\right)}{34020 a_3 \gamma  \omega ^2}r^5-\frac{1}{1306368 a_3^2 \gamma ^3 \omega ^3}\Bigg\{\gamma ^6-1212 a_3^2 \gamma ^3 \Lambda  \omega ^2-\nonumber\\
    &-216 a_3^4 \omega ^3 \left(3461 \gamma ^3+432 \Lambda ^2 \omega \right)-699840 a_3^6 \Lambda  \omega ^5+157464000 a_3^8 \omega ^6\Bigg\}r^6+\nonumber\\
    &+\frac{1}{1440270720 a_3^3 \gamma ^2 \omega ^4}\Bigg\{37 \gamma ^6-2325 a_3^2 \gamma ^3 \Lambda  \omega ^2108 a_3^4 \omega ^3 \left(243388 \gamma ^3+98901 \Lambda ^2 \omega \right)+\nonumber\\
    &+397124208 a_3^6 \Lambda  \omega ^5-69350294880 a_3^8 \omega ^6\Bigg\}r^7-\frac{1}{84652646400 a_3^4 \gamma ^4 \omega ^5}\Bigg\{7656 a_3^2 \gamma ^6 \Lambda  \omega ^2+\nonumber\\
    &+10497600 a_3^8 \omega ^6 \left(75475 \gamma ^3-8316 \Lambda ^2 \omega \right)+816480 a_3^6 \Lambda  \omega ^5 \left(1800 \Lambda ^2 \omega -14063 \gamma ^3\right)-\nonumber\\
    &-432 a_3^4 \omega ^3 \left(24494400 a_8 \gamma ^4 \omega ^2-52418 \gamma ^6+47397 \gamma ^3 \Lambda ^2 \omega -18900 \Lambda ^4 \omega ^2\right)+\nonumber\\
    &+3452240736000 a_3^{10} \Lambda  \omega ^8+127 \gamma ^9-36159245640000 a_3^{12} \omega ^9\Bigg\}r^8+\mathcal{O}(r^9).
\end{align}
Up to the twelfth order, all these solutions hold, but it is possible that at higher orders, some of these branches are non-consistent.
The splitting into different families is interesting but to see if the new branches are physical or at least, consistent in the context of quantum corrections to Einstein's equation a structural stability analysis is necessary.
 This could show if a small variation in \(\omega\) allows transitions in the different branches. 
\item
The $(2,2)$ solution.
\begin{align}
    &A(r)=a_2r^2+a_3 r^3+a_4 r^4+\mathcal{O}(r^5),\\
    &\frac{B(r)}{b_2}=r^2+b_3 r^3+\frac{1}{4} \left(\frac{a_3 b_3}{a_2}+4 a_2+b_3^2\right) r^4 +b_5r^5+\mathcal{O}(r^6),
\end{align}  
with five free parameters \((a_2,a_3,a_4,b_3,b_5)\) to this order.

This solution was checked to three higher orders, but the constraints on \(b_6(\gamma,\omega),b_5(\gamma,\omega)\) and \(b_7(\gamma,\omega)\) are too lengthy and do not give too much information on the solutions to be explicitly present. 
A fact that supports the consistency of this family is that it holds to orders where \((3,2)\) fails, while the general trend is that the higher \(n\) is, the longer the families take to break down.
\end{itemize}
Yet again, we checked, the $(n,2)$ family; with \(n\in \mathbb{N}/\{2\}\) and found they require \(\gamma\) to vanish. 
Similarly, The $(n,2\,+n)$ family; with \(n\in \mathbb{N}\), was tested for values \(n=\{0,1,2,3\}\) and again required \(\gamma\to 0\).

Thus, we can see that in this case, again, the only solutions that are consistent with the presence of the Einstein-Hilbert action are the two usual suspects \((0,0)\) and \((2,2)\).
\levelup{}
\dynsection{Summary of new solutions.}\label{sec:4.4}
Here we summarize the main novel power series solutions found.
\leveldown{}
\dynsection{\(\omega=0\).}
When studying quadratic gravity we found that for the \(\alpha=0\) case, new families of solutions, aside of the \((0,0)\), \((1,-1)\) and \((2,2)\), are possible. These correspond to, 
\begin{align}
    &(s= 1,t= -1),(s= 1,t= 0),(s= 2,t= -2),\nonumber\\
    &(s= 2,t= 2),(s= 7,t= -3),(s= 7,t= 8).
\end{align}
With the caveat that the last two families \((7,-3),\,(7,8)\), did not show dependence on \(\gamma\) at least for the first eight orders, therefore, we could not check their consistency. All we can say about these two families is that they satisfy the first eight orders of the equations of motion, but we are not yet ready to claim them to be consistent solutions for the established criterion.
\dynsection{\(\omega\neq0\).}
On the other hand, we have seen that the potentially unlimited families for the Weyl cubic action are, in presence of lower-order curvature operators, reduced to
\begin{equation}
    (0,0),\hspace{2.5cm}(2,2),
\end{equation}
which are the only possible solutions. What is more, both solutions appear to be present in every case where \(\omega\neq0\) including the \textit{two-loop Renormalization case}, studied in further detail in \cite{Alvarez2022}, 
\begin{equation}
    S=\int d^4 x \sqrt{|g|} \, \Big\{-\Lambda-\g \, R+\omega W_{\mu\nu\a\b}W^{\a\b\rho\sigma}W_{\rho\sigma}^{~~\mu\nu}\Big\}.
\end{equation}
Showing that a family of solutions is inconsistent with the EoMs, according to the convention established in \cref{sec:3.1}, might require going beyond eight orders. 
This seems to indicate that the families \((n,2)\) and \((n,2+n)\) might be solutions to the Weyl cubed action but the presence of the Einstein-Hilbert lagrangian is incompatible with them. This issue is not explored further as we are concerned by the quantum corrections to Einstein's equations for GR.
\levelup{}
\dynsection{Curvature singularities at \(r=0\).}\label{sec:4.5}
Spacetime singularities \cite{Clarke1994} are a subtle topic. Nevertheless, Penrose showed \cite{Penrose1965} that they are, under certain reasonable assumptions, unavoidable. 
The existence of a spacetime singularity is a concerning issue that some authors try to reconcile with General Relativity using the Cosmic Censorship Conjecture \cite{Penrose1969,Hawking2010}. 
Which postulates that under certain assumptions, singularities are to be hidden behind an Event Horizon.
All these are highly technical concepts studied in detail \cite{Carter2009, Carter2010} among many other classical referencces. The easier to grasp and more extended concept is that of a curvature singularity\footnote{Not all singularities are curvature singularities see \cite{Oliveira‐Neto}.}.

\textbf{Definition:}
\textit{A Manifold is said to have a \underline{curvature singularity at a point} \(\mathbf{p}\in \mathcal{M}\) if any of the scalar contractions of the curvature tensor or its contractions diverge at \(\mathbf{p}\).}

One can see that the three families \((1,-1)\), \((0,0)\) and \((2,2)\) satisfy that the Krestchman invariant given by; 
\begin{align}
    &\tensor{R}{_\alpha_\beta_\gamma_\delta}\tensor{R}{^\alpha^\beta^\gamma^\delta}=\frac{1}{4 r^4 A(r)^4 B(r)^4}\Bigg\{8 r^2 B(r)^4 A'(r)^2+16 (A(r)-1)^2 A(r)^2 B(r)^4+\nonumber\\
    &+r^4 \left(B(r) \left(A'(r) B'(r)-2 A(r) B''(r)\right)+A(r) B'(r)^2\right)^2+8 r^2 A(r)^2 B(r)^2 B'(r)^2\Bigg\},
\end{align}
satisfies,
\begin{itemize}
    \item If \(A(r)=1+a_2r^2+\cdots\) and \(B(r)=1+b_2r^2+\cdots\), then 
    \begin{equation}
        \lim_{r\to 0}\tensor{R}{_\alpha_\beta_\gamma_\delta}\tensor{R}{^\alpha^\beta^\gamma^\delta}=12 \left(a_2^2+b_2^2\right) \quad\mbox{finite}
    \end{equation}
    \item For Schwarzwschild, \(A(r)=r+a_2r^2+\cdots\) and \(B(r)=1/r+b_0+\cdots\),  
    \begin{equation}
        \lim_{r\to 0}\tensor{R}{_\alpha_\beta_\gamma_\delta}\tensor{R}{^\alpha^\beta^\gamma^\delta}\propto \frac{1}{r^6}\to \infty
    \end{equation}
    \item  Lastly, if \(A(r)=a_2r^2+a_4r^4+\cdots\) and \(B(r)=b_2r^2+b_4r^4\cdots\), then 
    \begin{equation}
        \lim_{r\to 0}\tensor{R}{_\alpha_\beta_\gamma_\delta}\tensor{R}{^\alpha^\beta^\gamma^\delta}\propto \frac{1}{r^8}\to \infty \label{sing2}
    \end{equation}
\end{itemize}
Then, from \cref{sing2} we can affirm that the \((2,2)\) solution has a curvature singularity at the origin. Furthermore, in \cite{Holdom2019} a numerical integration is used to show that the parameters of the \((2,2)\) family can be such that it has no Event Horizon and hence it corresponds to a naked singularity.

If the \((2,2)\) solution for the Weyl cubed operator was also shown to have parametric space corresponding to a naked singularity this would raise the relevance of the solution regarding gravitational collapse and the Cosmic Censorship cojecture. This is a really interesting topic that we leave for future task.

From our analysis one can see that there is legitimacy in raising the question of whether spherically symmetric and static solutions can remain horizonless in presence of quantum corrections or, more generally, on how do the solutions evolve when quantum perturbations are "turned on".
\levelup{}
\dynsection{Conclusions.}
In this thesis, we have performed a power series analysis to find static spherically symmetric solutions\footnote{In the sense of \cref{sec:D}.} to the EoMs of the  Einstein-Hilbert action with a Cosmological Constant and quadratic and cubic operators representing quantum corrections. 

First, for the case of quadratic gravity, we have reproduced the results from \cite{Stelle1977, Stelle1978, Stelle2015, Holdom2002, Alvarez2020}. 
Furthermore, we have explored in more detail the particular cases of quadratic gravity highlighted by Stelle and collaborators in \cite{Stelle2015} and later for a Cosmological Constant by \cite{Alvarez2020}. 
In this exploration we have found several new families that can be mapped to exact solutions of pure quadratic gravity \cite{Kehagias2015, Nelson2010, Whitt1985, Myung2013} and references therein.

Despite these interesting solutions, the most relevant feature of this work is, by far, the fact that the \((2,2)\) family, with a curvature singularity at the origin and potentialy\footnote{This was shown to be the case in quadratic gravity \cite{Holdom2019}, but it has not been shown in the presence of the Weyl cubic operator.} no Event Horizon, \cite{Holdom2002, Holdom2016, Holdom2019, Holdom2020, Stelle2015}, seems to be a solution when third-order curvature operators are present in the action.
The two-loop renormalization counterterm being compatible with the \((2,2)\) family but not with the Schwarzschild family \((1,-1)\) reinforces the physical significance of the former. 

This thesis has left several loose threads, to most obvious ones correspond to, 
\begin{itemize}
    \item Why does the number of free parameters increase with the order of the ODE considered? Does this point towards and additional symmetry of the action?
    \item Are the \((7,-3)\) and \((7,8)\) consistent solutions?
    \item What does the splitting of the \((0,0)\) family correspond to? Do all the non (Anti)de Sitter branches become incompatible at higher orders?
\end{itemize}
It remains to explore in detail whether the \((2,2)\) solution is compatible with solar system tests of General Relativity, whether it can be a result of stellar evolution in some situations, \cite{Holdom2002, Holdom2016, Holdom2019, Holdom2020}, and whether it presents any differential feature that could be used to differentiate it to the Schwarzschild \((1,-1)\) solution, including the posibility of it representing a Naked Singularity, which if realised could have an impact on the Cosmic Censorship conjecture.

The fact that the two-loop renormalization counterterm does not allow Schwarzschild-like solutions but does potentially admit \((2,2)\) solutions raises the question of whether the Schwarzschild deSitter solution which is the unique\footnote{Aside from Nariai in if \(\Lambda\neq0\).} spherically symmetric solution to the Einstein-Hilbert action, by Birkhoff's theorem \cite{Schleich2010}, is structurally stable or on the contrary, when the EoMs are modified by Weyl-cubic quantum corrections, the solutions are deformed to either the regular at the origin \cite{Stelle2015} \((0,0)\) family, or the singular at the origin \((2,2)\) family.
This is explored in some restricted cases on \cite{Alvarez2022}.

These results make the possibility of telling apart the  \((1,-1)\) and \((2,2)\) solutions crucial. Some numerical analyses on the issue have been made \cite{Holdom2016, Holdom2019, Holdom2020}. More work is necessary on this matter before any conclusions can be drawn.

We believe that this topic is full of diamonds waiting to be found as the most naive static and spherically symmetric case has shown many unforeseen features.

\dynsection{Acknowledgements.}
We would like to thank Enrique Álvarez Vázquez, for his insight, and patience. Also for always being ready to guide us on the physical aspects of any problem.

This work was produced while E.V. was financed by the "Ayudas para el fomento a la investigación en Estudios de máster UAM". 
The last revision of the manuscript took place while staying at the Centro de Ciencias Pedro Pascual in Benasque.
\clearpage
\appendix
\dynsection{Definitions, notation and conventions.}\label{sec:A}
In this appendix we summarize all conventions and definitions used throughout the work. 
\begin{itemize}
    \item For a space-time we use the definition of \cite{Hawking1973} but with the signature \((+,-,-,-)\).
    \item We will only work on the second order formalism, i.e., we assume the connection to be Levi-Civita connection; 
    \begin{equation}
        \tensor{\Gamma}{^\alpha_\mu_\nu}:=\begin{Bmatrix}
                                            \alpha \\
                                            \mu\;\nu
                                           \end{Bmatrix}=\frac{1}{2}\;\tensor{g}{^\alpha^\lambda}\left(\partial_\nu\tensor{g}{_\lambda_\mu}+\partial_\mu\tensor{g}{_\lambda_\nu}-\partial_\lambda\tensor{g}{_\mu_\nu}\right),\label{A.2}\tag{A.2}
    \end{equation}
    Recall the Levi-Civita connection is metric compatible, i.e., given the notion of covariant derivative defined by the connection, we have,
    \begin{equation}
        \nabla_\alpha \tensor{g}{_\mu_\nu}=0,\label{A.3}\tag{A.3}
    \end{equation}
    thus we can commute raising and lowering indices with covariant differentiation.
    \item The defintion for the Riemann curvature tensor is, 
    \begin{equation}
        \tensor{R}{^\mu_\alpha_\beta_\gamma}:=\partial_\beta\tensor{\Gamma}{^\mu_\alpha_\gamma}-\partial_\gamma\tensor{\Gamma}{^\mu_\alpha_\beta}+\tensor{\Gamma}{^\mu_\lambda_\beta}\tensor{\Gamma}{^\lambda_\alpha_\gamma}-\tensor{\Gamma}{^\mu_\lambda_\gamma}\tensor{\Gamma}{^\lambda_\alpha_\beta},\label{A.4}\tag{A.4}
    \end{equation}
    And, for the Ricci tensor and scalar,
    \begin{align}
        &\tensor{R}{_\alpha_\beta}:=\tensor{R}{^\mu_\alpha_\mu_\beta},\label{A.5}\tag{A.5}\\
        &R:=\tensor{R}{_\mu_\nu}\tensor{g}{^\mu^\nu}=\tensor{R}{_\mu^\mu}.\label{A.6}\tag{A.6}
    \end{align}
    \item With \eqref{A.5} and \eqref{A.6} we can define the Weyl tensor as, 
    \begin{equation}
        \tensor{C}{^\alpha_\beta_\gamma_\epsilon}:=\tensor{R}{^\alpha_\beta_\gamma_\epsilon}-\tensor{\delta}{^\alpha_\gamma}\tensor{K}{_\beta_\epsilon}-\tensor{\delta}{^\alpha_\epsilon}\tensor{K}{_\beta_\gamma}+\tensor{g}{_\beta_\gamma}\tensor{K}{_\epsilon^\alpha}-\tensor{g}{_\beta_\epsilon}\tensor{K}{_\gamma^\alpha},\label{A.7}\tag{A.7}
    \end{equation}
    where, \(\tensor{K}{_\alpha_\beta}\) is the Schouten tensor,
    \begin{equation}
        \tensor{K}{_\alpha_\beta}:=\frac{1}{n-2}\left(\tensor{R}{_\alpha_\beta}-\frac{1}{2(n-1)}R\tensor{g}{_\alpha_\beta}\right),\label{A.8}\tag{A.8}
    \end{equation}
    with \(n\) the number of dimensions, we will strictly consider \(n=4\).
\item \textbf{Gauss-Bonnet identity:} In 4 dimensions we have that the Euler characteristic of a manifold corresponds to, 
\begin{align}
    \chi(\mathcal{M})=\frac{1}{32\pi^2}\int d V\left(\tensor{R}{_\alpha_\beta_\gamma_\epsilon}\tensor{R}{^\alpha^\beta^\gamma^\epsilon}-4\tensor{R}{_\alpha_\beta}\tensor{R}{^\alpha^\beta}+R^2\right).\label{A.9}\tag{A.9}
\end{align}
\end{itemize}
Further properties of differential geometry are assumed familiar to the reader and can otherwise be checked in \cite{Hawking1973,Carroll2019} or each one's favourite textbook on the topic.
\clearpage
\dynsection{Equations of Motion.}\label{sec:B}
In this section, we present the explicit form of \cref{eom.1.1,eom.2.1}, that are solved using a Frobenius analysis in \cref{sec:4}.
To save some space we will drop the explicit \(r\) dependence of the functions \(A,\; B\) and their derivatives. Also, when useful, we will define auxiliary functions.
\begin{align}
        &H_{tt}=\frac{1}{144 r^6 A^7 B^5}\Bigg\{132 r^4 \omega  B^4 A'^4 \left(r B'-2 B\right)^2 -48 A^6 B^6 \left(3 \gamma  r^4+36 \beta  r^2-8 \omega \right)-\nonumber\\
        -&8 A^7 B^6 \left(9 \Lambda  r^6-18 \gamma  r^4+12 r^2 (\alpha -3 \beta )+4 \omega \right)+3 r^2 A^2 B^2\mathcal{F}\left(A,B,A',B',A'',B'',A^{(3)},B^{(3)}\right)+\nonumber\\
        +&A^4\mathcal{G}\left(A,B,A',B',A'',B'',A^{(3)},B^{(3)},B^{(4)}\right)-3 r A^3 B\mathcal{H}\left(A,B,A',B',A'',B'',A^{(3)},B^{(3)},B^{(4)}\right)+\nonumber\\
        +&r^3 \omega  A B^3 A'^2 \left(2 B-r B'\right) \Bigg[2 r B \left(63 r A'' B'+A' \left(213 r B''-97 B'\right)\right)-265 r^2 A' B'^2+\nonumber\\
        +&4 B^2 \left(149 A'-63 r A''\right)\Bigg]+6 A^5 B^2\mathcal{K}\left(A,B,A',B',A'',B'',A^{(3)},B^{(3)},B^{(4)}\right)\Bigg\}. \label{B.1.a}\tag{B.1.a}
\end{align}
Where we have defined, 
\begin{align}
    &\mathcal{F}\left(A,B,A',B',\cdots\right):=105 r^4 \omega  A'^2 B'^4-2 r^3 \omega  B A' B'^2 \left(27 r A'' B'+A' \left(130 r B''+23 B'\right)\right)+\nonumber\\
    +&8 r B^3 \Bigg[14 r A'^3 B' \left(r^2 (\alpha -3 \beta )+\omega \right)-2 r^2 \omega  A''^2 B'+\omega  A'^2 \left(r \left(31 B''-14 r B^{(3)}\right)-14 B'\right)+\nonumber\\
    +&r \omega  A' \left(\left(32 A''-2 r A^{(3)}\right) B'-23 r A'' B''\right)\Bigg]+4 r^2 \omega  B^2 \Bigg[r^2 A''^2 B'^2+\nonumber\\
    +&A'^2 \left(28 r^2 B''^2-60 B'^2+r B' \left(14 r B^{(3)}+43 B''\right)\right)+r A' B' \left(23 r A'' B''+\left(r A^{(3)}+13 A''\right) B'\right)\Bigg]-\nonumber\\
    -&16 B^4 \left(-r^2 \omega  A''^2+14 r A'^3 \left(r^2 (\alpha +6 \beta )+\omega \right)-20 \omega  A'^2+r \omega  A' \left(18 A''-r A^{(3)}\right)\right),\label{B.1.b}\tag{B.1.b}\\
    \nonumber\\
    &\mathcal{K}\left(A,B,A',B',\cdots\right):=8 B^4 \left(A' \left(3 \gamma  r^5+2 r \omega \right)+2 \left(r^2 (\alpha +15 \beta )-7 \omega \right)\right)+\nonumber\\
    +&49 r^4 B'^4 \left(r^2 (\alpha -3 \beta )+\omega \right)-16 r^4 B^3 \left(B^{(4)} \left(r^2 (\alpha -3 \beta )+\omega \right)+4 r (\alpha -3 \beta ) B^{(3)}\right)-\nonumber\\
    -&4 r^3 B B'^2 \left(29 r B'' \left(r^2 (\alpha -3 \beta )+\omega \right)+B' \left(r^2 (11 \alpha -78 \beta )-5 \omega \right)\right)+\nonumber\\
    +&4 r^2 B^2 \Bigg[9 r^2 B''^2 \left(r^2 (\alpha -3 \beta )+\omega \right)+B'^2 \left(3 \omega -r^2 (5 \alpha +12 \beta )\right)+\nonumber\\
    +&2 r B' \left(6 r B^{(3)} \left(r^2 (\alpha -3 \beta )+\omega \right)+B'' \left(r^2 (13 \alpha -66 \beta )-3 \omega \right)\right)\Bigg].\label{B.1.c}\tag{B.1.c}
\end{align}
And,
\begin{align}
    &\mathcal{G}\left(A,B,A',B',\cdots\right):=-4 r^3 B^3 \Bigg[B'^3 \left(40 \omega -87 r A' \left(r^2 (\alpha -3 \beta )+\omega \right)\right)+26 r^3 \omega  B''^3+\nonumber\\
    +&6 r^2 \omega  B' B'' \left(14 r B^{(3)}+B''\right)+6 r \omega  B'^2 \left(r \left(r B^{(4)}-6 B^{(3)}\right)-36 B''\right)\Bigg]+\nonumber\\
    +&24 r^2 B^4 \Bigg[-r B' \left(B'' \left(27 r A' \left(r^2 (\alpha -3 \beta )+\omega \right)-8 \omega \right)+2 r \omega  \left(r B^{(4)}+8 B^{(3)}\right)\right)-\nonumber\\
    -&2 B'^2 \left(3 r^2 A'' \left(r^2 (\alpha -3 \beta )+\omega \right)+A' \left(r^3 (6 \alpha -45 \beta )-3 r \omega \right)+2 \omega \right)+2 r^2 \omega  \Big((r B^{(3)})^2 -5 B''^2+\nonumber\\
    +&r \left(r B^{(4)}+2 B^{(3)}\right) B''\Big)\Bigg]+48 r^2 B^5 \Bigg[A' \Bigg(B' \left(r^2 (6 \beta -5 \alpha )+\omega \right)+r \Big(6 r B^{(3)} \left(r^2 (\alpha -3 \beta )+\omega \right)+\nonumber\\
    +&B'' \left(r^2 (13 \alpha -48 \beta )-\omega \right)\Big)\Big)+r \Bigg(2 r \left(2 A'' B'' \left(r^2 (\alpha -3 \beta )+\omega \right)+\omega  B^{(4)}\right)+\nonumber\\
    +&B' \left(r A^{(3)} \left(r^2 (\alpha -3 \beta )+\omega \right)+A'' \left(3 r^2 (\alpha -6 \beta )-\omega \right)\right)\Bigg)\Bigg]-32 B^6 \Bigg[3 \alpha  r^5 A^{(3)}+18 \beta  r^5 A^{(3)}+\nonumber\\
    +&3 r^3 \omega  A^{(3)}+3 A'' \left(r^4 (\alpha +6 \beta )-3 r^2 \omega \right)-6 A' \left(r^3 (\alpha +6 \beta )-4 r \omega \right)-10 \omega \Bigg]+121 r^6 \omega  B'^6-\nonumber\\
    -&6 r^5 \omega  B B'^4 \left(85 r B''-29 B'\right)+84 r^4 \omega  B^2 B'^2 \left(7 r^2 B''^2-4 B'^2+2 r B' \left(r B^{(3)}-2 B''\right)\right),\label{B.1.d}\tag{B.1.d}\\
\nonumber\\
    &\mathcal{H}\left(A,B,A',B',\cdots\right):=28 r^4 \omega  B B'^3 \left(r A'' B'-A' \left(B'-10 r B''\right)\right)-85 r^5 \omega  A' B'^5-\nonumber\\
    -&4 r^3 \omega  B^2 B' \Bigg[A' \left(55 r^2 B''^2-46 B'^2+2 r B' \left(10 r B^{(3)}+B''\right)\right)+r B' \left(18 r A'' B''+\left(r A^{(3)}+A''\right) B'\right)\Bigg]+\nonumber\\
    +&8 r B^4 \Bigg[r A'^2 \left(19 r B'' \left(r^2 (\alpha -3 \beta )+\omega \right)+B' \left(r^2 (13 \alpha -84 \beta )-5 \omega \right)\right)-2 r \omega  \Bigg(2 \left(A''-r A^{(3)}\right) B'+\nonumber\\
    +&r \left(r A^{(3)} B''+A'' \left(2 r B^{(3)}-5 B''\right)\right)\Bigg)+A' \Bigg(B' \left(13 r^2 A'' \left(r^2 (\alpha -3 \beta )+\omega \right)+4 \omega \right)-\nonumber\\
    -&2 r \omega  \left(2 B''+r \left(r B^{(4)}-8 B^{(3)}\right)\right)\Bigg)\Bigg]+2 r^2 B^3 \Bigg[-57 r A'^2 B'^2 \left(r^2 (\alpha -3 \beta )+\omega \right)+4 \omega  A' \Bigg(13 B'^2+\nonumber\\
    +&2 r^2 B'' \left(6 r B^{(3)}+5 B''\right)+B' \left(r^3 B^{(4)}-46 r B''\right)\Bigg)+4 r \omega  \Bigg(4 r^2 A'' B''^2-10 A'' B'^2+r B' \Big(r A^{(3)} B''+\nonumber\\
    +&A'' \left(2 r B^{(3)}+5 B''\right)\Big)\Bigg)\Bigg]+8 B^5 \Bigg[A'^2 \left(29 r \omega -r^3 (7 \alpha +60 \beta )\right)-4 r \omega  \left(r A^{(3)}-3 A''\right)-\nonumber\\
    -&2 A' \left(13 r^2 A'' \left(r^2 (\alpha +6 \beta )+\omega \right)+14 \omega \right)\Bigg].\label{B.1.e}\tag{B.1.e}
\end{align}

\begin{align}
    &H_{rr}=\frac{1}{144 r^6 A^5 B^6}\Bigg\{8 \left(9 \Lambda  r^6-18 \gamma  r^4+12 (\alpha -3 \beta ) r^2+4 \omega \right) A^6 B^6+\nonumber\\
    +&48 A^5 \left(\left(3 \gamma  r^4-36 \beta  r^2-4 \omega \right) B+r \left(3 \gamma  r^4+2 \omega \right) B'\right) B^5+11  \omega  (rA' \left(r B'-2 B\right) B)^3-\nonumber\\
    -&3 (r B)^2 \omega  A A' \left(r B'-2 B\right)^2 \Bigg[4 \left(5 A'-r A''\right) B^2+2 r \left(r B' A''+A' \left(6 r B''-4 B'\right)\right) B-7 r^2 A' B'^2\Bigg] -\nonumber\\
    -&6 A^4 B^2 \Bigg[16 \left(r^2 (\alpha -21 \beta )-3 \omega \right) B^4+16 {\Bigg( }r^3 \left(2 r (\alpha +6 \beta ) B''+\left((\alpha +6 \beta ) r^2+\omega \right) B^{(3)}\right)-\nonumber\\
    -2&\left(r^3 (\alpha +6 \beta )-r \omega \right) B'{\Bigg)} B^3+4 r^2 {\Bigg( }\left((\alpha -48 \beta ) r^2+\omega \right) B'^2-2 r \left(2 \left(3 (\alpha +3 \beta ) r^2+2 \omega \right) B''\right.+\nonumber\\
    +&\left.r \left((\alpha -3 \beta ) r^2+\omega \right) B^{(3)}\right) B'+r^2 \left((\alpha -3 \beta ) r^2+\omega \right) B''^2{\Bigg) } B^2+\nonumber\\
    +&4 r^3 B'^2 \left(\left((5 \alpha +12 \beta ) r^2+3 \omega \right) B'+3 r \left((\alpha -3 \beta ) r^2+\omega \right) B''\right) B-7 r^4 \left((\alpha -3 \beta ) r^2+\omega \right) B'^4\Bigg] B^2+\nonumber\\
    +&3 r A^2 B\Bigg[8  B^5\left(7 r \left((\alpha -12 \beta ) r^2+\omega \right) A'^2-8 \omega  A'+4 r \omega  A''\right)+2 r^2 \left(7 r \left((\alpha -3 \beta ) r^2+\omega \right) A'^2 B'^2+\right.\nonumber\\
    +&\left.8 r \omega  A'' \left(B'-r B''\right) B'+4 \omega  A' B^3 \left(3 B'^2+r \left(B''-2 r B^{(3)}\right) B'-2 r^2 B''^2\right)\right)+\nonumber\\
    -&8 r B^4\Big[7 r \left((\alpha +6 \beta ) r^2+\omega \right) B' A'^2-2 \omega  \left(4 B'+r \left(r B^{(3)}-3 B''\right)\right) A'-2 r \omega   A'' \left(r B''-3 B'\right)\Big]+\nonumber\\
    +&4 r^3 \omega  B' B^2\left(r B' A'' \left(B'+r B''\right)+A' \left(2 r^2 B''^2-5 B'^2+r \left(11 B''+r B^{(3)}\right) B'\right)\right) -\nonumber\\
    -&2 r^4 \omega  B'^3 \left(r B' A''+A'B \left(7 B'+9 r B''\right)\right) B+7 r^5 \omega  A' B'^5 \Bigg] +\nonumber\\
    +&A^3 \Bigg(11 \omega  B'^6 r^6-42 \omega  B B'^4 B'' r^6+12 \omega  B^2 B'^2 \left(-8 B'^2+r \left(5 B''+r B^{(3)}\right) B'+3 r^2 B''^2\right) r^4-\nonumber\\
    +&4 (r B)^3 \left(\left(16 \omega +9 r \left((\alpha -3 \beta ) r^2+\omega \right) A'\right) B'^3+48 r \omega  B'' B'^2-6 r^2 \omega  B'' \left(5 B''+r B^{(3)}\right) B'+2 r^3 \omega  B''^3\right) +\nonumber\\
    +&24 B^4 \Bigg(2 \omega  B'' \left(B''+r B^{(3)}\right) r^2-2 rB' \left(\left(5 \omega +r A' \left((\alpha -3 \beta ) r^2+\omega \right)\right) B''+2 r \omega  B^{(3)}\right) -\nonumber\\
    -&B'^2 \left(\left((\alpha -3 \beta ) r^2+\omega \right) A'' r^2+2 \left((2 \alpha +3 \beta ) r^2+\omega \right) A' r-2 \omega \right)\Bigg) r^2+\nonumber\\
    +&48 r B^5 \Bigg(2 \left(\left((\alpha +6 \beta ) r^2+\omega \right) A' B''+\omega  B^{(3)}\right) r^2+\nonumber\\
    +&B' \left(\left(r^3 (\alpha +24 \beta )-3 r \omega \right) A'+2 \left(\left((\alpha +6 \beta ) r^2+\omega \right) A'' r^2+\omega \right)\right)\Bigg)-\nonumber\\
    -&32 B^6 \left(3 \left((\alpha -12 \beta ) r^2+\omega \right) A'' r^2-6 \omega  A' r+4 \omega \right)\Bigg).\label{B.2} \tag{B.2}
\end{align}
\clearpage
\dynsection{Spherically symmetric and static metric.}\label{sec:D}
\textbf{Note:}\textit{The following appendix is just a compilation of the work in \cite{Hawking1973,Schleich2010} regarding static spherically symmetric spaces. 
It is not original but has great importance in setting the foundations for our investigation of spherically symmetric static spacetimes. }
\\
\\
The usual definition for spacetime to be spherically symmetric, \cite{Hawking1973, Schleich2010}, is based on the notion of a locally spherically symmetric tensor.

\textbf{Definition 1.-}\textit{A tensor field \(\tensor{T}{_a_b_\cdots ^c^d^\cdots}\), on a manifold \(\mathcal{M}^n\) \underline{localy spherically symmetric} if and only if every point in \(\mathcal{M}^n\) has an open neighbourhood \(U\subset \mathcal{M}^n\) such that the following conditions are satisfied:}
\begin{enumerate}
\item[(i)] \textit{There is a set of three independent vector fields \(\{\xi_i\}\), \(i=1,2,3\) on \(U\) which generate a faithful representation of the Lie algebra of \(SO(3)\) and the orbits of the vector fields are spacelike and two dimensional.}
\item[(ii)] \textit{\(\mathcal{L}_{\xi_i}\tensor{T}{_a_b_\cdots ^c^d^\cdots}\Big\vert_U=0 \) for these vector fields.}
\end{enumerate}

\textbf{Definition 2.-}\textit{A manifold \(\mathcal{M}^n\), with locally spherically symmetric metric \(\tensor{g}{_\alpha_\beta}\) is a \underline{locally} \underline{symmetric space}.}
\\
\\
That is, a spherically symmetric space admits \(SO(3)\) as a group of isometries, whose orbits are given by spacelike two-surfaces of constant positive curvature.

Then, if we choose a set of coordinates \(\{T,R,\theta,\phi\}\), such that the orbits are the surfaces \(T,R=\mbox{constant}\), while the orthogonal surfaces are \(\theta,\phi=\mbox{constant}\), then the metric must take the form, 
\begin{equation}
    \d s^2=\d\tau^2(T,R)-Y^2(T,R)\d \Omega_2^2(\theta,\phi)\label{D.1}\tag{D.1},
\end{equation}
with \(\d\tau^2\) some two surface, and \(\d \Omega_2^2\) the positive constant curvature two surface corresponding to the orbits.

Since \(\d\tau^2\) is an arbitrary two surface we can expand \cref{D.1} as, 
\begin{equation}
    \d s^2=D(T,R)\d T^2-2E(T,R)\d R\d T-C(T,R)\d R^2-Y^2(T,R)\d \Omega_2^2(\theta,\phi)\label{D.2}\tag{D.2},
\end{equation}
with the functions satisfying \(DC+E^2>0\) for the metric to have the right signature.

Then, with a change of coordinates we can write \eqref{D.2} as, 
\begin{equation}
    \d s^2=F(u,v)\d u^2-2X(u,v)\d u\d v -Y^2(u,v)\d \Omega_2^2(\theta,\phi)\label{D.3}\tag{D.3}.
\end{equation}
Now, we will make a particular choice on the general spherically symmetric metrics in \eqref{D.3}. 
We will only consider those that, by a suitable change of coordinates
\begin{equation}
    \{u,v\}\to\{t,r\},\label{D.4}\tag{D.4}
\end{equation}
can be taken into the form 
\begin{equation}
    \d s^2=B(r,t)\d t^2-A(r,t)\d r^2 -r^2\d \Omega_2^2(\theta,\phi)\label{D.5}\tag{D.5.1}.
\end{equation}
\cref{D.5} is what we call the \textit{Schwarzschild Gauge}, as the Schwarzschild solution is better known in such a gauge.
Actually, we will make the further simplifying assumption that the functions \(A(r,t),\;B(r,t)\) do not have explicit \(t\) dependence, i.e.,
\begin{equation}
    \d s^2=B(r)\d t^2-A(r)\d r^2 -r^2\d \Omega_2^2(\theta,\phi)\label{D.5b}\tag{D.5.2}.
\end{equation}

It is important to note that, when going from \cref{D.3} to \cref{D.5b} we are losing generality as there are spherically symmetric metrics that cannot be taken into the \textit{Schwarzschild Gauge}. An example of this is the Nariai metric \cite{Nariai1999,Nariai1999-2,Schleich2010} whose form in the \(\{t,r,\theta,\phi\}\) coordinate system is, 
\begin{equation}
    \d s^2= \left(1-\Lambda r^2\right)\d t^2-\frac{1}{\left(1-\Lambda r^2\right)}\d r^2-\frac{1}{\Lambda}\d \Omega_{D-2}.\label{D.6}\tag{D.6}
\end{equation}
Lastly, we will address the issue of the spacetime being \underline{static}.

\textbf{Definition 3.-}\textit{A manifold \(\mathcal{M}^n\),  is said to be \underline{static} if,}
\begin{enumerate}
    \item[(i)] \textit{It admits a \textbf{timelike} killing vector }
    \begin{equation}
    \mathcal{K}\rightarrow \mathcal{L}_\mathcal{K} \tensor{g}{_\alpha_\beta}=0 \;\&\;\mathcal{K}^2>0.\nonumber
    \end{equation}
        \item[(ii)] \textit{The killing vector field \(\mathcal{K}\), is orthogonal to a family of spacelike hypersurfaces.}
\end{enumerate}
It is clear that, in the \textit{Schwarzschild Gauge} \eqref{D.5b}, we have the killing vector,
\begin{equation}
    \mathcal{K}:=\partial_t\rightarrow\mathcal{K}^2=B(r)\label{D.7}\tag{D.7}
\end{equation}
however, if the spacetime has an horizon, the signature of the metric components \(A(r),\;B(r)\) might be inverted which translates into the vector \(\mathcal{K}\) changing its character from timelike to spacelike inside the horizon.

We will ignore the fact that in analytical extensions this change of sign might break condition (ii) for staticity and say, that spacetimes with a metric as in \cref{D.5b} are \textit{static and spherically symmetric spaces}, to mean that in the region outside the horizon these are spherically symmetric spaces that admit an additional timelike killing vector. 
Regardless of the fact that this vector field might change its nature to spacelike inside the horizon.

\newpage
\bibliography{bib}
\end{document}